\documentclass[numberedappendix,apj,twocolumn]{emulateapj}

\newcommand{\bFilter}{$B_{435}$}
\newcommand{\vFilter}{$V_{606}$}
\newcommand{\iFilter}{$i_{775}$}
\newcommand{\zFilter}{$z_{850}$}
\newcommand{\yFilter}{$Y_{105}$}

\newcommand{\jFilter}{$J_{125}$}

\newcommand{\hFilter}{$H_{160}$}

\shorttitle{The Bright End of the $z\sim8$ UV LF}
\shortauthors{Oesch et al.}

\begin{document}

\title{The Bright End of the UV Luminosity Function at $z\sim8$: \\New Constraints from CANDELS Data in GOODS-South
\altaffilmark{1}}

\altaffiltext{1}{Based on data obtained with the \textit{Hubble Space Telescope} operated by AURA, Inc. for NASA under contract NAS5-26555. }

\author{P. A. Oesch\altaffilmark{2,\dag},
R. J. Bouwens\altaffilmark{3}, 
G. D. Illingworth\altaffilmark{2}, 
V. Gonzalez\altaffilmark{2},
M. Trenti\altaffilmark{4}, \\
P. G. van Dokkum\altaffilmark{5},
M. Franx\altaffilmark{3}, 
I. Labb\'{e}\altaffilmark{3}, 
C. M. Carollo\altaffilmark{6}, 
D. Magee\altaffilmark{2}
}

\altaffiltext{2}{UCO/Lick Observatory, University of California, Santa Cruz, CA 95064; poesch@ucolick.org}
\altaffiltext{3}{Leiden Observatory, Leiden University, NL-2300 RA Leiden, Netherlands}
\altaffiltext{4}{University of Colorado, Center for Astrophysics and Space Astronomy,389-UCB, Boulder, CO 80309, USA}
\altaffiltext{5}{Department of Astronomy, Yale University, New Haven, CT 06520}
\altaffiltext{6}{Institute for Astronomy, ETH Zurich, 8092 Zurich, Switzerland}
\altaffiltext{\dag}{Hubble Fellow}

\begin{abstract}
We present new $z\sim8$ galaxy candidates from a search over $\sim$95 arcmin$^2$ of WFC3/IR data, tripling the previous search area for bright $z\sim8$ galaxies. Our analysis uses newly acquired WFC3/IR imaging data from the CANDELS Multi-Cycle Treasury program over the GOODS South field.
These new data are combined with existing deep optical ACS imaging to
search for relatively bright ($M_{UV} < -19.5$ mag) $z\sim8$ galaxy candidates
using the Lyman Break technique.
These new candidates are used
to determine the bright end of the UV luminosity function (LF) of
star-forming galaxies at $z\sim7.2-8.7$, i.e. a cosmic age of $600\pm80$ Myr. 
To minimize contamination from lower redshift galaxies, we make full use of all optical ACS data and impose strict non-detection criteria based on an optical $\chi^2_{opt}$ flux measurement. In the whole search area we identify 16 candidate $z\sim8$ galaxies, spanning a magnitude range $H_{160,AB} =25.7-27.9$ mag.  The new data show that the UV LF is a factor $\sim1.7$ lower at $M_{UV} < -19.5$ mag than determined from the HUDF09 and ERS data alone. Combining this new sample with the previous candidates from the HUDF09 and ERS data allows us to perform the most accurate measurement of the $z\sim8$ UV LF yet. Schechter function fits to the combined data result in a best-fit characteristic magnitude of $M_*(z=8) = -20.04\pm0.46$ mag. The faint-end slope is very steep, though quite uncertain, with $\alpha=-2.06\pm0.32$.  A combination of wide area data with additional ultra-deep imaging will be required to significantly reduce the uncertainties on these parameters in the future.

\end{abstract}


\section{Introduction}

Thanks to the unprecedented efficiency of the WFC3/IR camera on-board the \textit{Hubble Space Telescope} (\textit{HST}), the last two years have seen a remarkable progress in the exploration of galaxies within the first Gyr after the Big Bang. Immediately after its installation, WFC3/IR has been used to obtain the deepest NIR images ever seen as part of the HUDF09 program \citep[PI: Illingworth; e.g.][]{Bouwens11c}. This pushed the observational frontier of galaxies from $z\sim6$ well into the reionization epoch, with first constraints even at $z\sim10$ \citep{Bouwens11a,Oesch11,Zheng12}. 

In combination with deep ancillary optical data from the HUDF \citep{Beckwith06} and GOODS surveys \citep{Giavalisco04a}, the WFC3/IR data from the HUDF09 and the ERS \citep{Windhorst11} programs allowed for the identification of more than 130 $z\sim7-8$ galaxy candidates to date using the Lyman Break Galaxy (LBG) selection technique \citep[e.g.][]{Steidel96,Giavalisco02}. Since the Ly$\alpha$ absorption of the neutral inter-galactic medium shifts to $>1$\micron\ at $z\gtrsim7$, such galaxies can only be detected in the NIR data and are completely invisible in the optical. For example, $z\sim8$ galaxies can be selected as $Y_{105}$-dropouts, based on their very red $Y_{105}-J_{125}$ colors and optical non-detections.

The first $z\sim7-8$ galaxy candidates have already been used to measure the UV LF at these redshifts \citep[e.g.][]{Oesch10a,Bouwens10a,Bouwens11c,McLure10,Bunker10,Finkelstein10,Yan10,Wilkins10,Lorenzoni11,Bradley12}. Additionally, these candidates allowed for first estimates of the physical parameters of $z>6$ galaxies, such as their sizes \citep{Oesch10b}, UV continuum slopes \citep[e.g.][]{Bouwens10b,Bouwens11d,Finkelstein10,Finkelstein11,Dunlop11}, rest-frame optical colors \citep[][]{Gonzalez11}, and even ages and stellar masses \citep[e.g.][]{Labbe10a,Labbe10b,Gonzalez10a,Gonzalez10b,Finkelstein10,Schaerer10}. Furthermore,  $z\sim8$ proto-cluster candidates have been identified in pure parallel WFC3/IR data \citep{Trenti11b}.

While the first deep WFC3/IR data sets provide good constraints on the faint population at $z>7$, they only probe a limited volume and therefore result in poor constraints on the much less abundant bright galaxies around the exponential cut-off of the LF. At $z\sim7$, constraints on this bright cut-off can be obtained using ground-based, wide-area data from, e.g., Subaru or the VLT \citep[e.g.][]{Ouchi09,Castellano10a,Hickey10}. However, beyond $z\sim7$, the galaxy population is too faint, and the sky is too bright to efficiently detect $z\gtrsim8$ galaxies from the ground with current facilities, resulting only in upper limits at $M_{UV}< -22$ mag \citep{Castellano10b} in the field. Note that ground-based searches behind lensing clusters resulted in a few potential $z\geq8$ candidates \citep[see e.g.][]{Laporte11}, and in limits on the bright end of the $z\sim9$ LF \citep{Laporte12}.

Due to the limited volume probed at $z\sim8$ with current deep \textit{HST} data, the bright end of the $z\sim8$ LF has thus remained quite uncertain, subject to large cosmic variance. For example, in the HUDF09-2 field data \citet{Bouwens11c} identified two very bright ($H_{160,AB} \sim 26$ mag) sources \citep[see also][]{Wilkins11a,McLure11}, which possibly biased the previous measurement of the $z\sim8$ UV LF towards somewhat higher values. It is therefore essential to study larger area WFC3/IR data to quantify how representative these bright candidates are of the $z\sim8$ galaxy population.

There are a few ongoing and planned \textit{HST} programs that can be used for this task. First, the Multi-Cycle Treasury program CANDELS is obtaining $Y_{105}$ data over the two GOODS fields \citep{Giavalisco04a}, resulting in a total search area for $z\sim8$ galaxies of $\sim300$ arcmin$^2$. Second, the pure parallel programs BORG \citep{Trenti11a,Trenti11b} and HIPPIES \citep[][]{Yan11}, are designed to find rare bright $z\sim8$ galaxy candidates, sampling a combined total of about $400$ arcmin$^2$ to varying depths. Before the advent of JWST, the combination of different WFC3/IR data sets probing different scales will likely be the only way to obtain a well-sampled UV LF at $z\sim8$ \citep[see e.g.][]{Bradley12}.

In this paper, we take advantage of the full $Y_{105}$-band data obtained over the GOODS-South field as part of the CANDELS program to identify relatively bright $z\sim8$ galaxy candidates at absolute magnitudes $M_{UV}< -19.5$ mag. We combine these with our previous $z\sim8$ searches in the ultra-deep data from the HUDF09 as well as the ERS programs to significantly improve the sampling of the bright end of the $z\sim8$ UV LF. With the inclusion of the CANDELS field, we essentially triple the search area for bright $z\sim8$ galaxies in GOODS-South relative to our previous analysis, resulting in much more reliable constraints on the shape of the bright end of the LF and the luminosity at which an exponential cutoff might occur.

An accurate measurement of the bright part of the UV LF is critical for enabling meaningful comparisons with galaxy evolution models at these redshifts \citep[e.g.][]{Finlator11a,Finlator11b,Trenti10,Lacey11,Jaacks11,Dayal11,Forero10,Salvaterra11,Munoz11}. The bright end potentially contains important information on the star-formation efficiencies, formation timescales and duty cycles of galaxies within dark matter halos in the first few hundred Myr of cosmic time. 

Furthermore, a well sampled bright end of the LF helps to break degeneracies in fitting the Schechter function \citep{Schechter76}. Even if shallower data sets do not provide real information on the faint-end slopes, they can still decrease the uncertainties on the faint-end slope, assuming that the shape of the LF can accurately be described by a Schechter function \citep[e.g., see discussion in][]{Bouwens08}. Accurate estimates of the faint-end slope are critical for obtaining realistic constraints on the number of ionizing photons emitted by the ultra-faint galaxy population at these redshifts. Within the current uncertainties it is unclear whether the faint galaxy population was luminous enough to reionize the universe alone at $z\gtrsim7$ \citep[e.g.][]{Bolton07,Oesch09,Robertson10b,Bouwens11b,Shull11}.

Finally, the identification of brighter $z\sim8$ galaxy candidates is crucially important to provide candidates for spectroscopic follow-up. Spectroscopy has proven to be extremely challenging, mainly due to the intrinsic faintness of the sources, the abundance of bright night-sky lines, and the absorption of a significant fraction of the Ly$\alpha$ flux of these galaxies by the neutral inter-galactic medium (IGM). Therefore the progress in spectroscopic follow-up has been relatively slow, with only a handful of confirmed sources at $z\sim7$ \citep[see e.g.][]{Schenker11,Pentericci11,Ono11}.
At $z\sim8$, possibly the best chance for spectroscopic confirmation of the redshifts of candidates is with upcoming multi-object NIR spectrographs targeting simultaneously a few bright sources identified in contiguous wider area imaging.

This paper is organized as follows. After the presentation of the data in Section \ref{sec:data} we explain our source selection criteria  and present the $z\sim8$ candidates in Section \ref{sec:selection}. These are then used to derive constraints on the bright end of the $z\sim8$ UV LF in Section \ref{sec:LFconstraints}.
 We will refer to the HST filters F435W, F606W, F775W, F814W, F850LP, F105W, F125W, F160W as \bFilter, \vFilter, \iFilter, $I_{814}$, \zFilter, \yFilter, \jFilter, \hFilter, respectively.
Throughout this paper, we adopt $\Omega_M=0.3, \Omega_\Lambda=0.7, H_0=70$ km~s$^{-1}$Mpc$^{-1}$, i.e. $h=0.7$ \citep[WMAP-7;][]{Komatsu11}. Magnitudes are given in the AB system \citep{Oke83}.

\begin{deluxetable*}{lccccccccc}
\tablecaption{$5\sigma$ Depth and Area of Data Used in this Paper\label{tab:data}}
\tablewidth{0 pt}
\tablecolumns{9}
\tablehead{\colhead{Field} & Area [arcmin$^2$]\tablenotemark{$\dagger$} & \bFilter &\vFilter &\colhead{\iFilter} &  $I_{814}$\tablenotemark{*} &  \colhead{\zFilter}  &\colhead{$Y_{105}$} & \colhead{\jFilter} & \colhead{\hFilter}   }

\startdata
CANDELS-Deep & 60.2 &  27.9  & 28.1  & 27.6  & 28.5  & 27.6  & 27.9  & 28.0  & 27.7 \\
CANDELS-Wide & 34.2 & 27.9 & 28.1 & 27.6 & 28.0 & 27.6  & 27.2  & 27.3  & 27.0 

\enddata

\tablecomments{The limits correspond to $5\sigma$ variations in the sky flux measured in
a circular aperture of 0\farcs25 radius, i.e. no correction to total magnitude was performed. The full CANDELS-Deep is about 3.5 orbits per filter, and CANDELS-Wide one orbit per filter, leading to a difference of about 0.7 mag. }
\tablenotetext{$\dagger$}{The quoted area corresponds to the part of the image where imaging in all three WFC3/IR filters is available as well as ACS in $B_{435}$, $V_{606}$, $i_{775}$, $z_{850}$.}
\tablenotetext{*}{$I_{814}$ is used for confirming the optical non-detection of candidates using a 2$\sigma$ non-detection criterion. All candidates listed in table \ref{tab:phot} do have $I_{814}$ coverage.}
\end{deluxetable*}

\begin{figure}[tbp]
	\centering
	
	\includegraphics[scale=5]{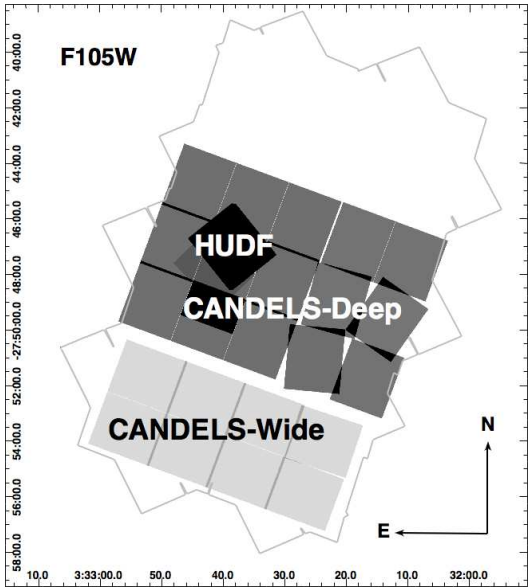}
  \caption{The WFC3/IR F105W data over the GOODS South field that are used in this analysis. These data include all WFC3/IR data taken over this area before the end of 2011. 
The different fields are color coded by exposure time, with deeper areas being darker. The CANDELS-Deep covers $\sim65$ arcmin$^2$ with $\sim$3.5 orbits of F105W data. The outline indicates the area of the optical ACS coverage of GOODS.
The CANDELS-Wide field spans 34 arcmin$^2$ and is only of 1 orbit depth.
The central part covered by the HUDF is omitted from our primary analysis since it was analyzed fully by \citet{Bouwens11c}. However, we do combine the CANDELS $z\sim8$ candidates with the ones we previously selected in the deep HUDF09 and the ERS fields \citep[see][]{Bouwens11c} when constraining the $z\sim8$ LF. 
  For more information on the data used here see Section \ref{sec:data} and Table \ref{tab:data}.  }
	\label{fig:fields}
\end{figure}

\begin{figure*}[tbp]
	\centering
	\includegraphics[scale=0.5]{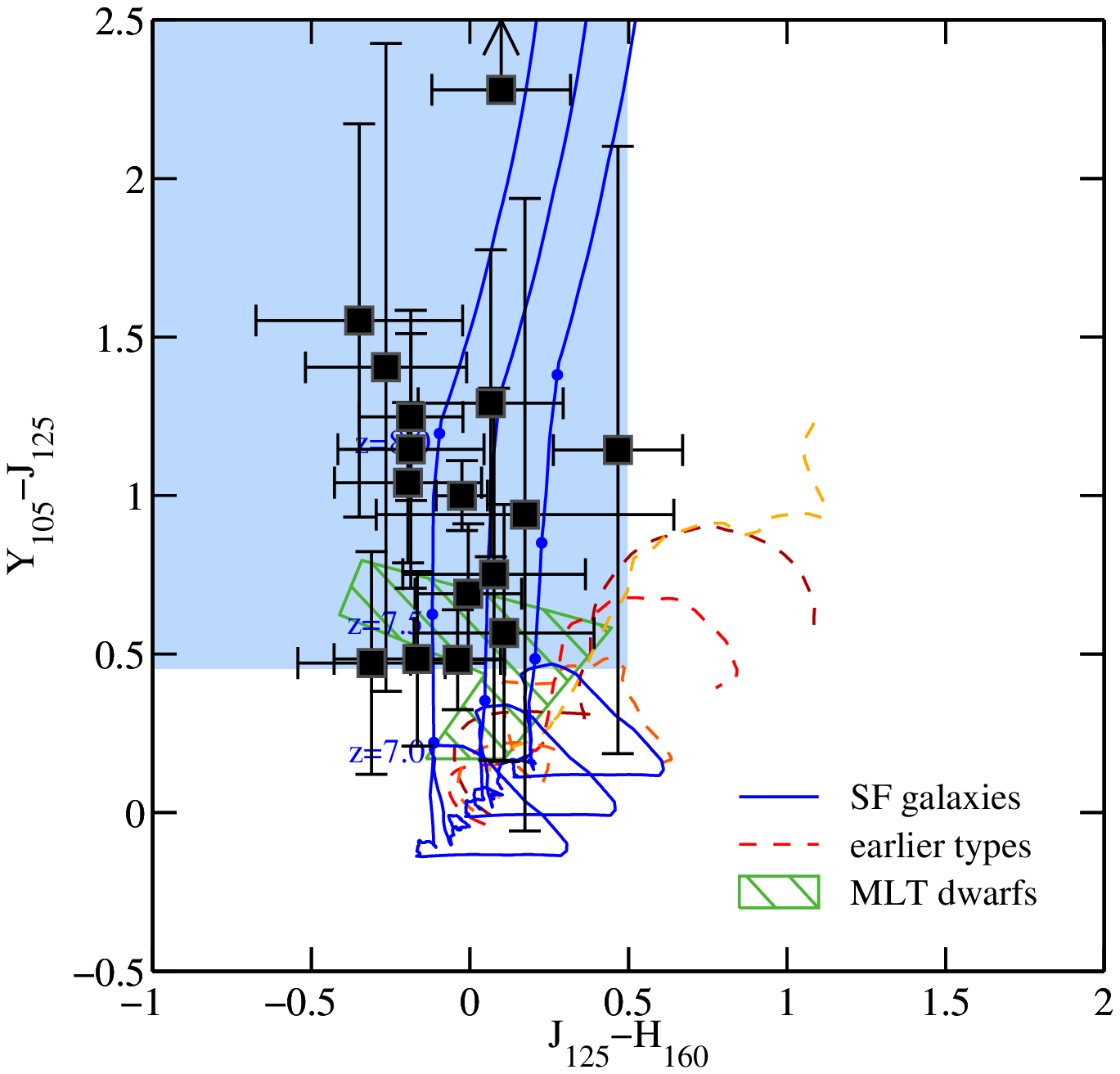} 
	\includegraphics[scale=0.5]{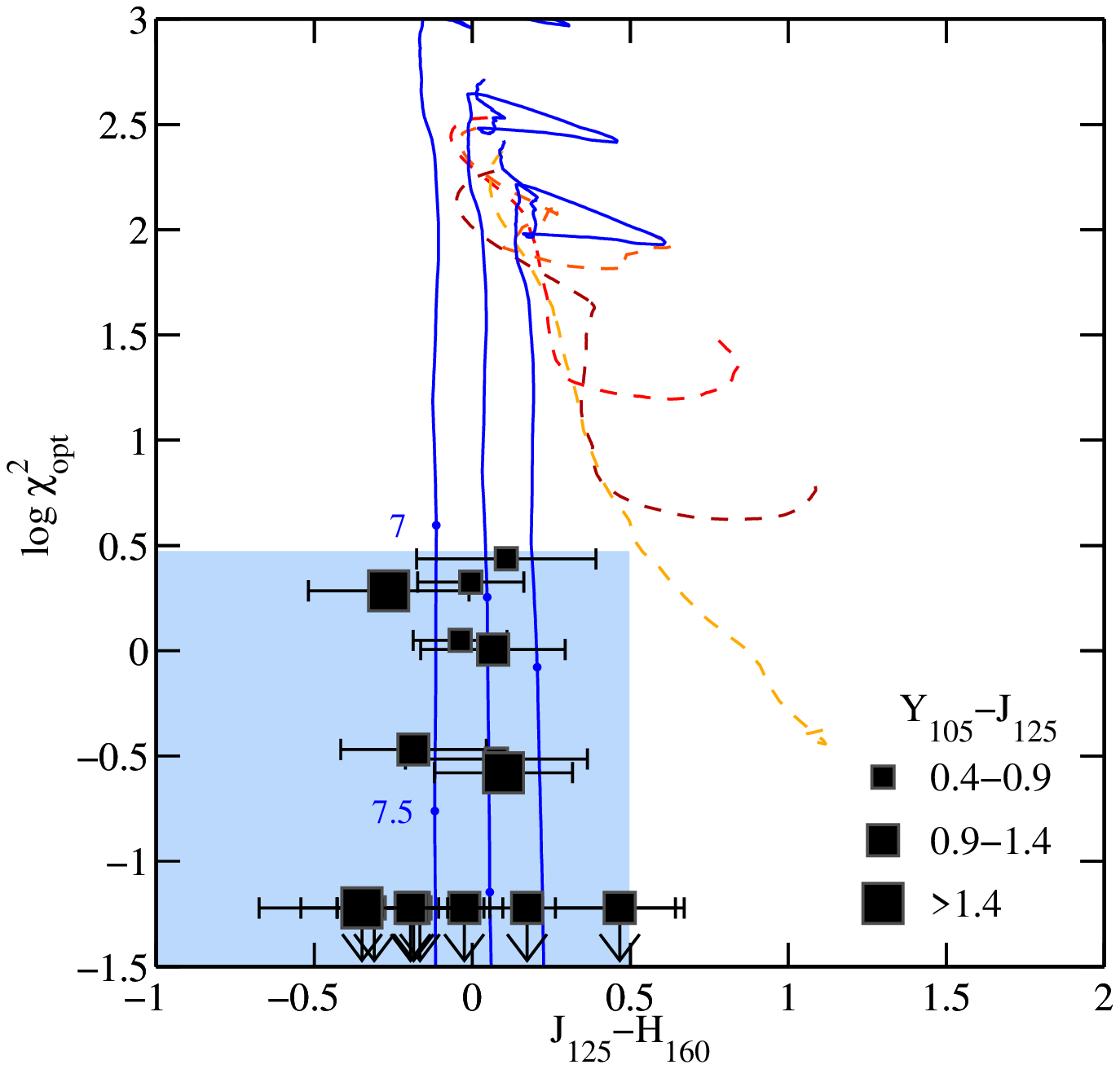} 
  \caption{\textit{Left -- } The color selection criterion used to identify $z\sim8$ galaxies. The color selection (blue shaded area) is identical to the one used by \citet[][]{Bouwens11c}. This ensures that the resulting galaxy samples can be
readily combined in the analysis. Additionally, we impose strict non-detection requirements based on the optical $\chi^2$ values (see text). This enables us to reliably select galaxies to quite blue $Y_{105}-J_{125}$ colors.
  The redshifted color tracks of different types of galaxies are shown as solid (star-forming) and dashed lines (earlier types). The latter are shown up to $z=3$, while the tracks of star-forming galaxies are extended to $z>8$. Their locations at $z=7, ~7.5,$ and $8$ are labelled and marked with small dots. The three blue lines correspond to different amounts of dust reddening $E(B-V)=0, ~0.1, ~0.2$ (left to right). The expected location of ultra-cool M, L, and T dwarf stars is indicated by a green hatched region. The colors of the $z\sim8$ candidates identified in this paper are shown as black squares with the usual $1\sigma$ error bars. 
  \textit{ Right -- } The optical $\chi^2$ for galaxies with $H_{160,AB} = 27.0$ mag against $J_{125}-H_{160}$ color. This constraint is particularly effective in excluding  intermediate redshift galaxies (see Section \ref{sec:chi2opt} and Figure \ref{fig:chi2opt}). 
   The lines correspond to the same galaxy types and redshift ranges as in the left panel. The measurements for the galaxy candidates identified in the CANDELS data are shown as black squares. Upper limits are shown for galaxies with $\log\chi^2_{opt}<-1.2$. The size of the plot symbols represent the measured $Y_{105}-J_{125}$ colors. Only one source with $Y_{105}-J_{125}<0.9$ lies near to the track of passive, intermediate redshift galaxies in this $\chi^2_{opt}$ vs. color diagram.  }
	\label{fig:colcol}
\end{figure*}

\section{The Data}
\label{sec:data}

\subsection{WFC3/IR Data over GOODS-South}

In this paper, we analyze the complete public WFC3/IR data obtained over the GOODS South field as part of the Multi-Cycle Treasury program CANDELS \citep[PI: Faber/Ferguson;][]{Grogin11,Koekemoer11} and then combine it with the deeper HUDF09 and ERS datasets.
The completion of the $Y_{105}$ imaging over the CANDELS GOODS South field (see Figure \ref{fig:fields}) has made this a unique data set for studying the bright end of the $z\sim8$ UV LF based on a $Y_{105}$-dropout selection.

The CANDELS data is split in two parts, CANDELS-Deep and CANDELS-Wide. The first covers the central part of GOODS South in $3\times5$ tiles with $\sim$3.5 orbits in each filter $Y_{105}$, \jFilter\ and \hFilter. These data cover $\sim65$ arcmin$^2$, reaching down to $H_{160,AB} = 27.7$ mag.
The CANDELS-Wide program consists of data in the same three WFC3/IR filters as for the CANDELS-Deep. These were fully acquired already by end of March, 2011 and cover 8 WFC3/IR pointings ($\sim35$ arcmin$^2$) to $\sim1$ orbit depth in each filter, thus reaching to $H_{160,AB}=27.0$ mag.

In addition, we combined all the public WFC3/IR data that have been taken over these fields before March 2012 from other programs. In particular, we included the imaging data of the supernovae follow-up program of CANDELS (PI: Riess), which adds imaging over a few pointings of CANDELS Deep. The final exposure map of all the \yFilter\ data included is shown in Figure \ref{fig:fields}. The central part covered by the ultra-deep HUDF09 program was omitted from our analysis of the CANDELS data in order not to replicate the bright candidates of that field given in the study of \citet{Bouwens11c}. We will use those candidates, however, when computing the total $z\sim8$ LF in section \ref{sec:z8LF} (see also section \ref{sec:deepdata}).

The WFC3/IR data was reduced using standard techniques. 
 The pipeline-processed science frames were obtained from MAST, and were subsequently registered to the existing optical ACS data at a pixel scale of $0\farcs06$. All input images were
inspected visually for satellite trails and other artefacts, such as loss
of the guide star.
Pixels affected by persistence from previous
observations, as identified from the persistence masks provided by
STScI\footnote{http://archive.stsci.edu/prepds/persist/}, have been flagged
and removed from our reductions.
For the last few visits of F105W data, we found that the cosmic ray subtraction of the archived data was not satisfactory. We therefore masked all pixels that were affected by cosmic rays in the
archived data which resulted in much cleaner reductions.

The optical ACS data used here include the original GOODS optical imaging as well as additional data available over these fields from supernova follow-up programs. These reach about $0.1-0.3$ mag deeper in $z_{850}$ than the v2.0 reductions of GOODS. This is critical for excluding interlopers to the $z\sim8$ galaxy selections. We restricted our analysis to the part of the WFC3/IR data where imaging in all three filters (as well as full ACS coverage) was available. This results in 60.2 arcmin$^2$ and 34.2 arcmin$^2$ for CANDELS Deep and Wide, respectively. 

Finally, we also include a full reduction of all the ACS F814W filter data available over GOODS South. Such data was mainly taken in parallel to other observations of several programs. In particular, most of the CANDELS WFC3/IR observations obtained F814W data in parallel. However, over the years, several such programs have been conducted, and the resulting data reach to a non-uniform depth of $I_{814}\sim28.0-28.5$ mag, i.e. they are deeper than the original GOODS $i_{775}$ data, and in several parts of the field are essentially the deepest filter data now.
The specifics of all the data used in this paper are summarized in Table \ref{tab:data}.

\subsection{Deeper WFC3/IR Data}
\label{sec:deepdata}

For accurately constraining the UV LF at $z\sim8$ it is crucial to probe the galaxy population over as large a dynamic range in luminosity as possible. In the later sections of this paper, when we compute the LF, we therefore include all the $z\sim8$ galaxy candidates that we previously identified in \citet{Bouwens11c}. These candidates were selected from the three ultra-deep WFC3/IR pointings of the HUDF09 program (PI: Illingworth) as well as from  $\sim$40 arcmin$^2$ of WFC3/IR Early Release Science data \citep[ERS; see][]{Windhorst11}. In total, these are 59 $z\sim8$ galaxy candidates with magnitudes in the range $H_{160,AB} = 26.0-29.4$ mag over the total area of $\sim53$ arcmin$^2$. For more information on these candidates and their selection, we refer the reader to \citet{Bouwens11c}. As we describe below, we use essentially identical
selection procedures to \citet{Bouwens11c}, to ensure that we can
combine the new LF results and the older ones without corrections or
biases.

\section{Source Selection}
\label{sec:selection}

\subsection{Catalog Construction} 
Source catalogs are obtained with the SExtractor program \citep{Bertin96}, which is used to detect galaxies in the square-root of chi-square image \citep{Szalay99} computed from the $J_{125}$ and $H_{160}$ data. We perform matched aperture photometry on point spread function (PSF)-matched images. The colors used here are based on small elliptical apertures (1.2 Kron; these are typically nearly round and about 0.2\arcsec\ long axis), and total magnitudes are measured in standard 2.5 Kron apertures (typically 0.4\arcsec\ radius), corrected to total fluxes using the PSF encircled energy of an equivalent aperture to account for flux loss in PSF wings (typically $\sim0.15-0.2$ mag).

The input RMS maps were scaled to properly represent the flux variations in $0\farcs25$ radius apertures, which were used to establish the detection significance of sources. The RMS scaling was done based on the detected variation in 1000 random apertures per WFC3/IR frame on empty sky regions after $3\sigma$ clipping. This procedure ensures that the weight maps correctly reproduce the actual noise in the images. Subsequently only sources with signal-to-noise ratios (S/N) larger than 4.5 in $H_{160}$  are considered (as measured in $0\farcs25$ radius circular apertures).

\subsection{$z\sim8$ Color-Color Selection}
\label{sec:candsel}

Thanks to the strong IGM absorption in the rest-frame UV, $z\sim8$ galaxies can be selected in broad-band color-color diagrams. At $z\gtrsim7$ the Ly$\alpha$ absorption shifts into the $Y_{105}$ band, rendering galaxies red in $Y_{105}-J_{125}$, while the $J_{125}-H_{160}$ color of star-forming galaxies is still expected to be blue. The selection is discussed here, but the most
challenging aspect is to minimize the contamination from low redshift
objects, as we discuss in Sections \ref{sec:chi2opt} and \ref{sec:contamination}.

We adopt the same color selection criteria for our $z\sim8$ selection as in \citet{Bouwens11c}, which will allow us to directly combine the previous data with the candidates identified here. Specifically, the selection criteria are:
\begin{eqnarray*}
	(Y_{105}-J_{125})&>&0.45 \\ 	
	(J_{125}-H_{160})&<&0.5 
\end{eqnarray*}
Additionally, we require sources to be detected at 4.5$\sigma$ in both $J_{125}$ and $H_{160}$, and at 5$\sigma$ in at least one of these two bands.
Finally, we require candidates to be detected at $<2\sigma$ in all the optical ACS data, and also require candidates to meet an optical $\chi^2_{opt}$ flux constraint. Both tests play a major role in removing potential low-redshift contaminants (see Section \ref{sec:chi2opt}). 
Finally, we checked the $I_{814}$ images of all sources and additionally required galaxies to be $2\sigma$ non-detections also in $I_{814}$. This last test helped to eliminate three sources which are likely contaminants. 

The above color  criterion is illustrated in Figure \ref{fig:colcol}, where we show the expected colors of star-forming galaxies using 100 Myr old stellar population models from \citet{Bruzual03}. These are additionally reddened with small amounts of dust using the dust law of \citet{Calzetti00}. The IGM absorption is modeled using \citet{Madau95}. Our color criteria start to select such galaxies $z\gtrsim7.2$. The selection then peaks around $z\sim8$ and falls off due to the combination of an increased distance modulus
and flux reduction as the Lyman Break approaches the filter bandpass
limit at higher redshift \citep[see Fig. 4 in][]{Bouwens11c}. The mean redshift of our sample is $\langle z \rangle = 7.9$, with 80\% of galaxies expected to lie at $z=7.2-8.7$ (see Fig \ref{fig:selfun}).
 
In Figure \ref{fig:colcol}, we also show the colors of lower redshift galaxies using \citet{Coleman80} templates, as well as cool dwarf stars \citep{Burgasser04a}, which could potentially contaminate our selection. Despite some overlap in this color-color diagram with $z>7$ galaxy candidates, we expect the contamination of such sources to be small due to our strict optical non-detection criteria (see next section and discussion in section \ref{sec:contamination}). 
This is illustrated in the right panel of Figure \ref{fig:colcol}, where we show the expected $\chi^2_{opt}$ values (defined in the next section) as a function of $J_{125}-H_{160}$ color for different galaxy types. Most intermediate redshift galaxies are expected to be well detected in the optical data, resulting in $\log\chi^2_{opt}>1$. Only completely quiescent galaxies would lie close to our selection box. The figure also includes the measured values of the CANDELS $z\sim8$ candidates (see section \ref{sec:z8candidates}). The sizes of the symbols represent their $Y_{105}-J_{125}$ colors. Our selection includes only three galaxies with relatively blue $Y_{105}-J_{125}$ colors with positive $\chi^2_{opt}$ values between 1 and 3. These would have the highest probability of being lower redshift contaminants.

As a further step against interlopers, we check each candidate for bright detections in the \textit{Spitzer} IRAC 3.6\micron\ images available over GOODS-South \citep[e.g.][]{Dickinson03}. This has proven to be a very effective test for removing dusty contaminants in very high-z galaxy searches in which only limited information on the UV-continuum slope of galaxies is available \citep[see e.g.][]{Oesch11}. Young, star-forming galaxies at $z>7$ with small amounts of dust reddening are expected to show colors $H_{160}-[3.6]\lesssim1$ \citep[see also][]{Gonzalez11}. Any galaxy with substantially redder colors is therefore likely to be a lower redshift contaminant. 

Indeed, in constructing our sample, we identified one bright ($H_{160,AB} = 24.3$ mag) edge-on spiral galaxy (at 03:32:14.72,$-$27:46:21.6), which satisfied our WFC3/IR color and optical non-detection criteria. However, this source is extremely luminous in the IRAC data, with $H_{160}-[3.6]=2.3$, and is also clearly detected even in the [8.0] band, as expected for a heavily dust obscured source at intermediate redshift.  We therefore removed it from the potential $z\sim8$ galaxy sample. 
All other sources where we could measure IRAC fluxes (i.e. which were not completely blended with bright foreground sources) were consistent with the limit $H_{160}-[3.6]<1$.

\begin{figure}[tbp]
	\centering
	\includegraphics[width=\linewidth]{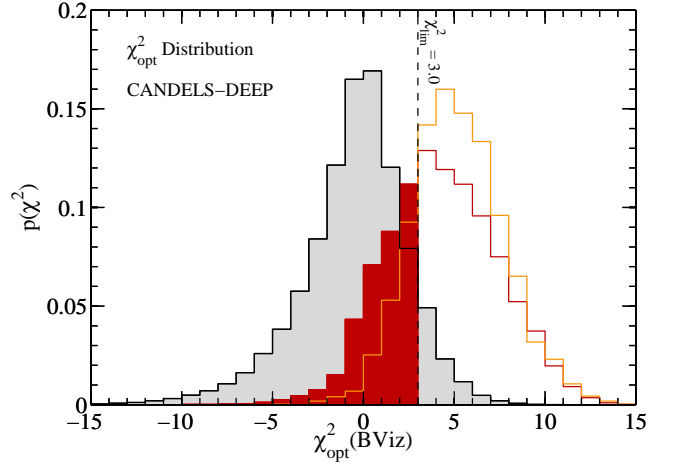} 
  \caption{The optical $\chi^2$ non-detection criterion in the CANDELS Deep data. The histograms show the different distribution functions for empty sky positions (gray filled), as well as what is expected for contaminants based on our simulations. True $z\sim8$ sources will have a distribution
like the gray empty sky distribution. The orange distribution is based on simulations using galaxies from the HUDF09 to which we applied Gaussian flux scatter appropriate to our $\sim2$ mag shallower CANDELS data. The dark red histogram corresponds to simulations using brighter galaxies in the CANDELS field itself, which are dimmed and have the appropriate flux scatter applied. The distributions from the two simulations are very similar. The dark red filled histogram represents the fraction of all contaminants that satisfy the color and non-detection criteria, but which lie below the adopted $\chi^2_{opt}$ limit of $3.0$ (see text), and which would be confused with
real $z\sim8$ objects. With this cut we are able to reduce the contamination rate by a factor $\sim3$-$4$ (the ratio of the dark red area to
the total). The residual contamination in our fields is expected to be about 1 source in the CANDELS Deep data and $<0.1$ in CANDELS Wide giving a total contamination rate of about 10\%. }
	\label{fig:chi2opt}
\end{figure}

\begin{deluxetable*}{lccccccccl}
\tablecaption{Table of $z\sim8$ Galaxy Candidates in the CANDELS GOODS-South Data*\label{tab:phot}}
\tablewidth{0 pt}
\tablecolumns{8}

\tablehead{\colhead{ID} & $\alpha$ & $\delta$ &\colhead{$H_{160}$}  &\colhead{$J_{125}-H_{160}$} & \colhead{$Y_{105}-J_{125}$} &  \colhead{S/N ($H_{160} / J_{125}$)}  & \colhead{Alternate ID\tablenotemark{a} } }

\startdata

\cutinhead{CANDELS-Deep}

 CANDY-2499448181  &  03:32:49.94  &  $-$27:48:18.1  &  $25.70\pm0.09$ & $-0.0\pm0.1$ & $1.0\pm0.1$  &  22.0/30.0  & ISO\_085   \\    
 CANDY-2402644099  &  03:32:40.26  &  $-$27:44:09.9  &  $26.19\pm0.16$ & $-0.2\pm0.3$ & $0.5\pm0.3$  &  4.6/6.4      &  \\ 
 CANDY-2468950074  &  03:32:46.89  &  $-$27:50:07.4  &  $26.20\pm0.14$ & $-0.0\pm0.1$ & $0.5\pm0.2$  &  11.0/12.2  &  \\
 CANDY-2209751370  &  03:32:20.97  &  $-$27:51:37.0  &  $26.46\pm0.14$ & $-0.2\pm0.2$ & $1.2\pm0.3$  &  7.8/8.9      & AUTO\_212 \\ 
 CANDY-2272447364  &  03:32:27.24  &  $-$27:47:36.4  &  $26.70\pm0.20$ & $-0.2\pm0.2$ & $1.0\pm0.3$  &  6.1/8.3      &   \\ 
 CANDY-2320345371  &  03:32:32.03  &  $-$27:45:37.1  &  $26.74\pm0.26$ & $-0.0\pm0.2$ & $0.7\pm0.2$  &  13.1/13.0  &   \\ 
 CANDY-2350049216  &  03:32:35.00  &  $-$27:49:21.6  &  $26.90\pm0.16$ & $0.1\pm0.2$ & $>2.3$  &  7.1/7.7                 & ISO\_063\\  
 CANDY-2139147577  &  03:32:13.91  &  $-$27:47:57.7  &  $27.09\pm0.28$ & $0.1\pm0.3$ & $0.6\pm0.4$  &  6.2/5.0        & \\
 CANDY-2243349150  &  03:32:24.33  &  $-$27:49:15.0  &  $27.11\pm0.20$ & $-0.3\pm0.2$ & $0.5\pm0.4$  &  5.9/9.2      & \\
 CANDY-2181952456  &  03:32:18.19  &  $-$27:52:45.6  &  $27.14\pm0.19$ & $-0.2\pm0.2$ & $1.1\pm0.4$  &  7.0/8.9      & AUTO\_204 \\
 CANDY-2209848535  &  03:32:20.98  &  $-$27:48:53.5  &  $27.22\pm0.28$ & $0.1\pm0.2$ & $1.3\pm0.5$  &  8.2/8.4        & ISO\_071 \\ 
 CANDY-2209246371  &  03:32:20.92  &  $-$27:46:37.1  &  $27.31\pm0.25$ & $0.1\pm0.3$ & $0.8\pm0.6$  &  5.8/6.5       & \\ 
 CANDY-2277945141  &  03:32:27.79  &  $-$27:45:14.1  &  $27.76\pm0.35$ & $-0.3\pm0.3$ & $1.6\pm0.6$  &  4.5/6.5     & \\ 
 CANDY-2432246169  &  03:32:43.22  &  $-$27:46:16.9  &  $27.88\pm0.43$ & $0.2\pm0.5$ & $0.9\pm1.0$  &  4.5/5.0       &   \\

\cutinhead{CANDELS-Wide}

CANDY-2379552208  &  03:32:37.95  &  -27:52:20.8  &  $26.45\pm0.12$ & $0.5\pm0.2$ & $1.1\pm1.0$  &  9.7/7.7   & \\ 
CANDY-2408551569  &  03:32:40.85  &  -27:51:56.9  &  $27.30\pm0.22$ & $-0.3\pm0.3$ & $1.4\pm1.0$  &  5.4/8.8  & 

\enddata

\tablenotetext{*}{ ~Limits are 1$\sigma$.}
\tablenotetext{a}{IDs refer to Yan et al., ApJ submitted, eprint arXiv:1112.6406v2 (see also appendix).}
\end{deluxetable*}

\subsection{Optical Non-detections Using $\chi^2_{opt}$}
\label{sec:chi2opt}

As indicated above, one of the main challenges in selecting robust LBGs is to remove intermediate redshift contaminants. Given the high efficiency of WFC3/IR the available ancillary optical data in most of the WFC3/IR search fields is not appreciably deeper than the new IR data (see e.g. Table \ref{tab:data}). Thus at the faintest magnitudes even a $2\sigma$-nondetection criterion, which is the standard that is used in LBG selections, is not sufficiently effective to eliminate faint, dusty interlopers. 

In previous papers \citep[e.g.][]{Bouwens11c}, we have developed an efficient method for eliminating low-$z$ contaminants by making full use of all the information in the optical data. In particular, for each galaxy we compute an optical pseudo $\chi^2_{opt}$ value from its aperture flux measurements of all optical bands as
\[
\chi^2_{opt} = \sum_i \mathrm{SGN}(f_i) (f_i/\sigma_i)^2
\] 
where $i$ runs over \bFilter\, \vFilter\, \iFilter\, and \zFilter, and SGN is the sign function, i.e. $\mathrm{SGN}(x) = -1$ if $x<0$ and $\mathrm{SGN}(x) = 1$ if $x>0$.

For real high-redshift candidates the $\chi^2_{opt}$ distribution is expected to be centered around zero, while for contaminants the distribution is skewed toward positive values. Therefore, by removing galaxies with values above some limiting value $\chi^2_{lim}$ it is possible to significantly reduce the number of contaminants in LBG samples.

The adopted limiting value is derived from two different sets of simulations. The results are shown in Figure \ref{fig:chi2opt} where we compare the
observed $\chi^2_{opt}$ to those from our simulations to establish the limiting value.

The first simulation is based on the HUDF09 data, where the available optical and WFC3/IR data are $\sim1.5-2$ mag deeper than in the fields studied here. At a given $H_{160,AB}$ magnitude, this allows us to check what fraction of sources would contaminate our sample, if they were observed at the shallower depth of our data. We use a Monte-Carlo simulation in which we apply Gaussian scatter to the fluxes of the HUDF09 sources scaled to the depth of the CANDELS Deep and Wide data. We then compute the $\chi^2_{opt}$ values for all sources which did not satisfy our selection criteria in the ultra-deep data, but which would have been selected and would have passed the optical $2\sigma$ non-detection criterion. The distribution of $\chi^2_{opt}$ values of these contaminants is shown in Figure \ref{fig:chi2opt} as an orange histogram.

For the second simulation, we use brighter sources directly from the CANDELS fields which are dimmed to fainter magnitudes. Thanks to the wider area relative to the HUDF, this allows us to probe rarer interloper populations and so is complementary to the HUDF09 simulation. We only use sources with $H_{160,AB} = 24-25.5$ mag as input, in order to keep the population as close as possible to the magnitude range of our candidates. Again, for each source satisfying the color selection after applying Gaussian noise, we compute the value of $\chi^2_{opt}$. This is shown in Figure \ref{fig:chi2opt} as a dark red line.

As is apparent from the Figure, the $\chi^2_{opt}$ distribution of interlopers is significantly skewed to positive values relative to regions of empty sky as expected for real $z\sim8$ sources. After some
consideration we adopted a limit $\chi^2_{opt} < 3.0$ as being a good discriminator
between $z\sim8$ galaxies and contaminating sources.
By using this limit of $\chi^2_{opt} < 3.0$, it is possible to reduce the contamination rate by a factor of $\sim3-4$ relative to a simple $2\sigma$ non-detection criterion in all optical bands \citep[see also Appendix D of][]{Bouwens11c}. At the same time, we lose just $\sim$15\% of real $z\sim8$ sources.  We correct for this small
loss in the estimate of the selection volume.

The expected  residual number of contaminants, after applying the $\chi^2_{opt}$ limit, is $1.0\pm0.5$ sources in the CANDELS Deep data, and $<0.1$ in the whole CANDELS Wide field. Based on these simulations, we thus estimate a contamination rate that is only $\sim10\%$ for our full sample.
Note that, given our ignorance of the LF of interlopers and cosmic variance in the input galaxy sample for our photometric scatter simulations, it is conceivable that the true contamination rate is as high as $15-20$\%.

\begin{figure*}[tbp]
	\centering
	\includegraphics[scale=0.7]{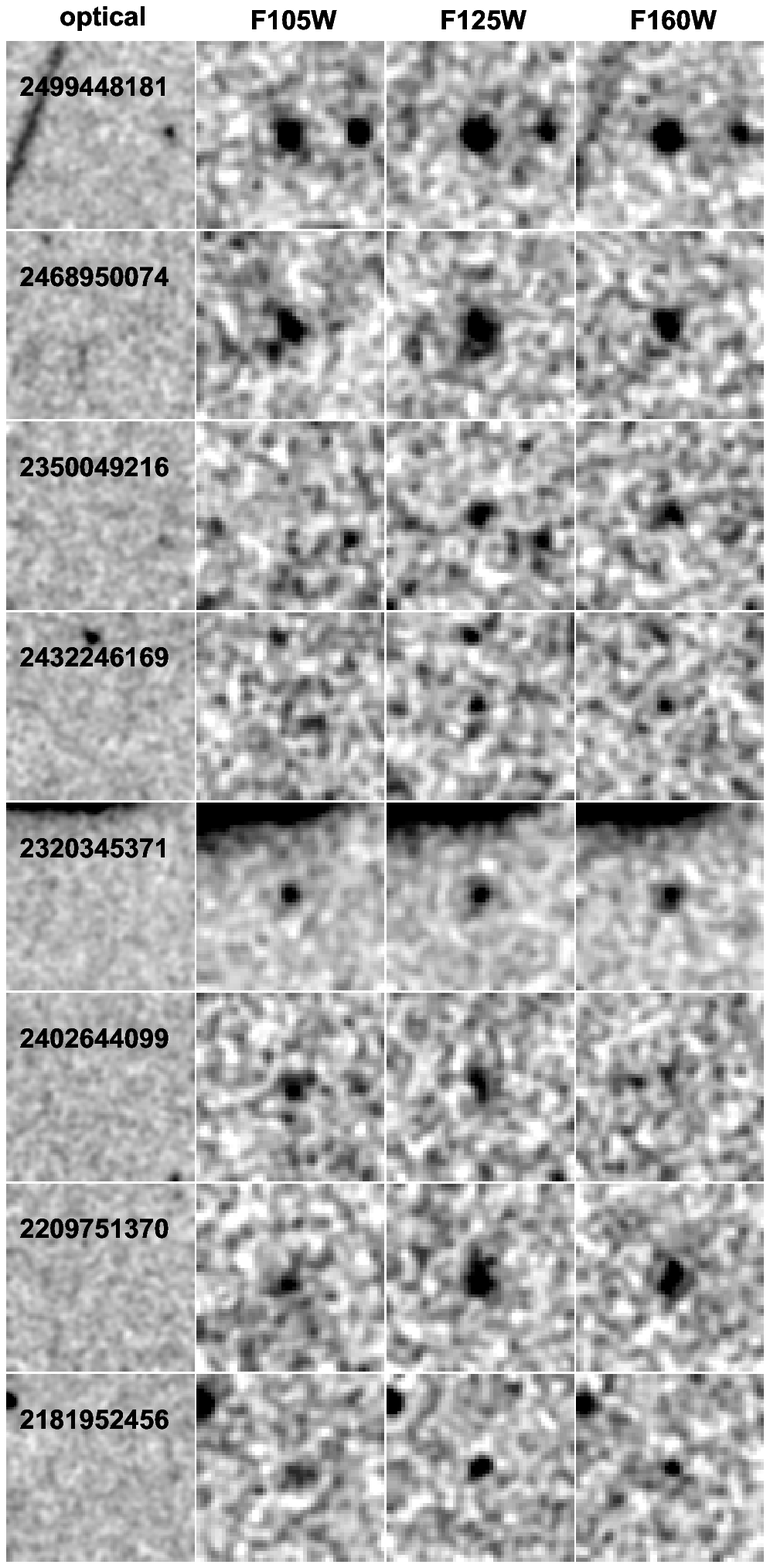}  
    \includegraphics[scale=0.7]{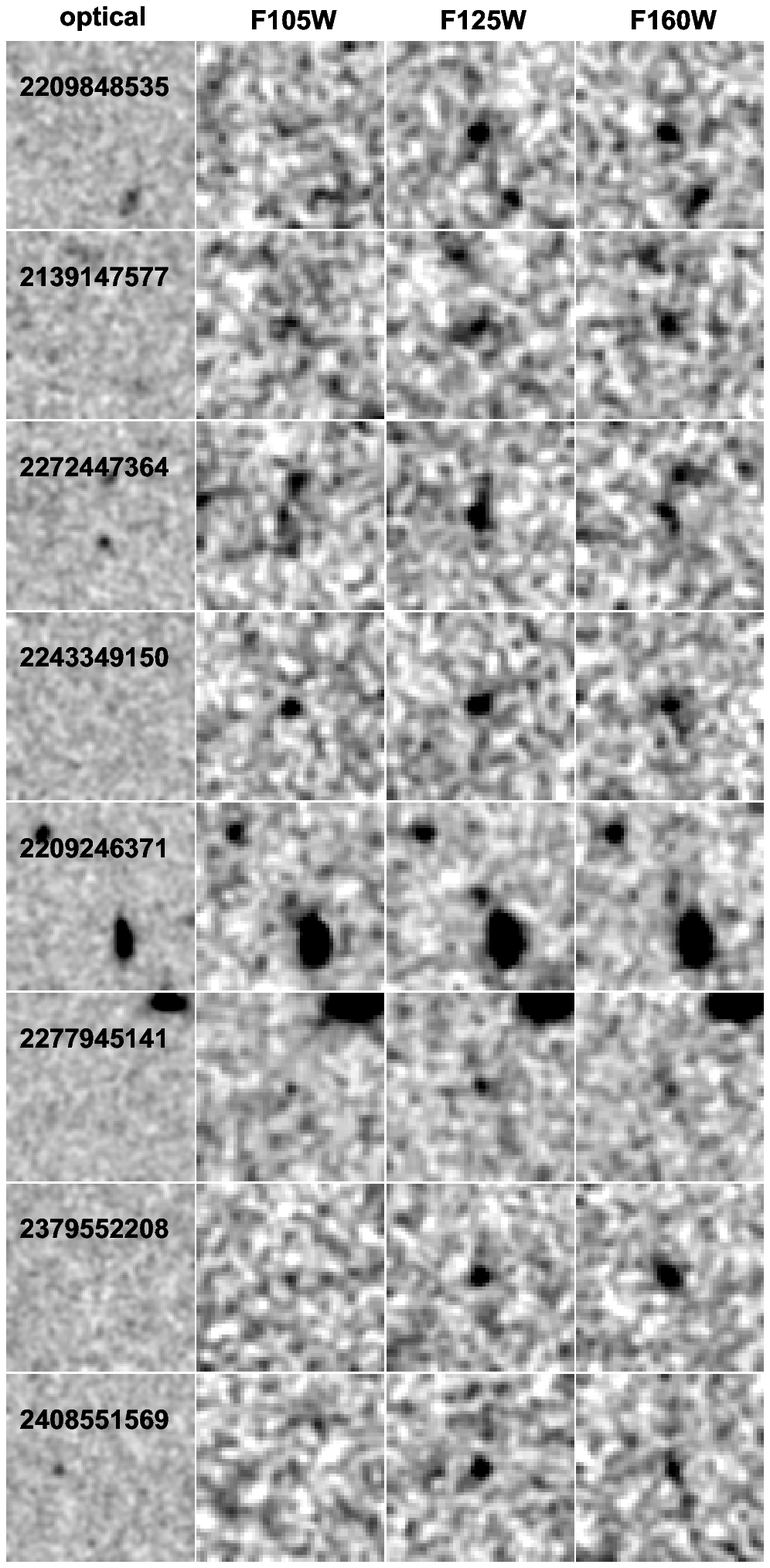} 
  \caption{ Images of the $z\sim8$ galaxy candidates identified in the CANDELS data. From left to right, these show: (1) a stack of all the optical ACS data (BViz),   (2) WFC3/IR data in the filters $Y_{105}$, (3) $J_{125}$, and (4) $H_{160}$. The images are 3 arcsec on a side, oriented North up, East left. All candidates are well detected in both $J_{125}$ and $H_{160}$, show a significant flux decrement to $Y_{105}$, and show no flux in the optical. An unsmoothed version of these images are shown in the appendix in Figure \ref{fig:stampssm}. }
	\label{fig:stamps}
\end{figure*}

\subsection{The Bright CANDELS $z\sim8$ Candidates}
\label{sec:z8candidates}

Applying all the selection criteria outlined above, we identify 16 new $z\sim8$ galaxy candidates in the full CANDELS $Y_{105}$ data over the GOODS-South field; 14 in CANDELS Deep, and two in CANDELS Wide. Their properties are listed in Table \ref{tab:phot}, and individual images for all candidates in the WFC3/IR bands as well as a stack of all the optical ACS images are shown in Figure \ref{fig:stamps}. 

The candidates span a range of $H_{160,AB} = 25.7-27.9$ mag. Our brightest source is thus among the most luminous $z\sim8$ candidates known to date. It is $\sim0.5$ mag brighter than all other CANDELS sources, and would be an ideal target for future spectroscopic follow-up. 

Such a luminous object raises an interesting possibility. If we assume that the UV luminosity of galaxies scales with halo mass, such a bright $z\sim8$ galaxy is expected to lie in an over-dense environment with $\sim5$ fainter galaxies within a diameter of 60 arcsec \citep[see e.g.][]{Trenti11b,Munoz08}. Unfortunately, the source lies close to an extremely bright star, reducing the detectability of fainter companions significantly. Nevertheless, we would have expected to see a few fainter companions around this source. None are found, however. Unfortunately, the proximity to the bright star also makes it impossible to obtain IRAC flux measurements from the current \textit{Spitzer} data to strengthen or refute the high redshift solution of this source. Deeper optical or IR data (or a spectrum) may thus be needed to fully resolve the origin of our brightest candidate. 
It completely satisfies our
selection criteria and our $\chi^2_{opt}$ limit such that it is most likely at $z\sim8$,
and so we just note the caveat raised by the clustering estimates.

 As can be seen from Figure \ref{fig:stamps}, all $z\sim8$ candidates are quite compact, but appear to be resolved. Their size distribution is consistent with the expectations based on extrapolation
from lower redshift samples out to $z\sim7$ \citep[e.g.][]{Oesch10b}. 
Only two sources show signs for elongated morphologies, indicative of disk structures. 
However, the surface brightness limit in the CANDELS data work against clearly 
detecting such structures, compared to e.g. the HUDF \citep[see e.g.][]{Oesch10b}.


At the same time our paper was first submitted, the CANDELS team also published a catalog of $z\sim8$ galaxy candidates from the DEEP area in \citet{Yan12a}. A significant fraction of their sources (45\%) does not satisfy our strict optical non-detection criteria, with a few showing clear optical detections, thus disqualifying them for being at $z>7$. All their candidates are discussed in detail in the appendix, where we also tabulate our measurements for these sources.

\subsection{Sample Contamination}
\label{sec:contamination}

Thanks to our restrictive non-detection criteria we do not expect high contamination levels in our samples. Nevertheless, some residual contamination can not be excluded. In particular, we briefly discuss several potential sources of contamination below:

1. \textit{ -- Low Mass Stars:} As shown in Figure \ref{fig:colcol}, ultra-cool dwarf stars can exhibit similarly red colors as the potential high-redshift galaxies. Given the resolution of WFC3/IR, however, we can directly check whether any of our sources are unresolved. This test can reliably be done only for the
brighter candidates, with well sampled profiles. Indeed, we did remove one source from our candidate list with colors consistent with a T-dwarf which appeared to be point-like (at 03:32:25.33,  $-$27:48:54.2). 
Furthermore, the contamination by dwarf stars is also expected to be low based on their observed surface densities in high galactic latitude fields \citep[e.g.][]{Ryan05,Ryan11}.

2. \textit{ -- Photometric Scatter: }  The most probable source of contamination in our sample comes from photometric scatter of sources with intrinsic colors similar to high-$z$ galaxies. This has been quantified already in section \ref{sec:chi2opt}. To summarize, after applying the $\chi^2_{opt}<3.0$ cut, we only expect $1.0\pm0.5$ contaminants due to photometric scatter in the CANDELS Deep data set, and $<0.1$ in the CANDELS Wide. Therefore, with our strict optical non-detection criteria, we limit the amount of contamination to $\sim10\%$. 

Similar conclusions are reached by investigating the $\chi^2_{opt}$ distribution. As can be seen in the right panel of Figure \ref{fig:colcol}, our sample includes five galaxies with $\chi^2_{opt}>0.9$. Statistically, we expect only $22\%$ of galaxies above such values, i.e. 3.5 sources. This would suggest again a contamination of $\sim10\%$ in the full sample.

3. \textit{ -- Spurious Sources: } All our candidates are very well detected both in $H_{160,AB}$ as well as in $J_{125}$. The chance for spurious $>4.5\sigma$ detections at the same location in both bands is negligibly small. Additionally, most of the sources are detected at lower significance in $Y_{105}$, further reducing the chance of a spurious source. Spurious sources are clearly not a concern for our sample.

4. \textit{ -- Transients: } Since the optical data has been taken a few years prior to the new WFC3/IR data, it is possible that supernovae that went off in the meantime are selected as infrared detections without any optical counterparts. However, since one of the specific goals of CANDELS-Deep is to search for such supernovae, the \hFilter\ and \jFilter\ data acquisition has been distributed over several different epochs. So far, eight epochs have been acquired. This allows us to check directly whether any of our sources could potentially be a supernova. We therefore group the individual epochs in four bins sorted by exposure date and check whether all our sources are still visible  in these sub-splits (in a $H_{160} + J_{125}$ image). All CANDELS-Deep candidates are indeed detected in these images. For the two candidates in CANDELS-Wide data, this test is not as decisive, as the  $J_{125}$ and $H_{160}$ data acquisition has only been split over two epochs. Furthermore, we stress again that these sources do not appear to be point-like. We therefore conclude that supernovae have not contaminated our sample.

\begin{figure}[tbp]
	\centering
	\includegraphics[scale=0.5]{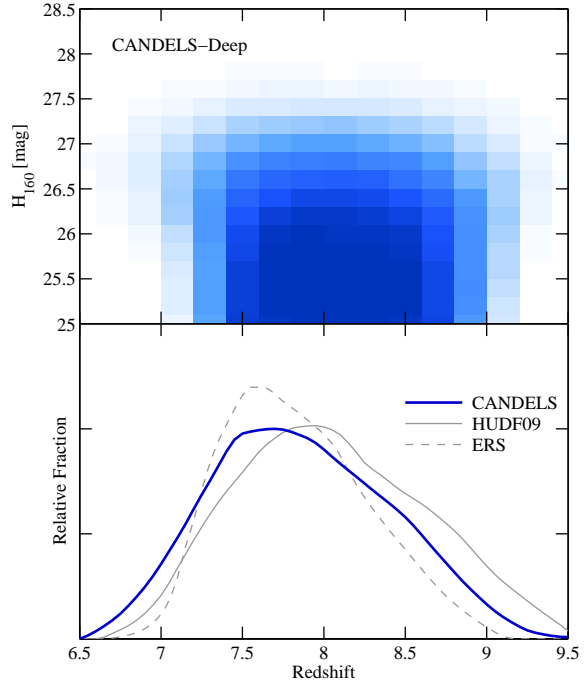} 
  \caption{ \textit{Top -- } The magnitude and redshift dependent selection function of the CANDELS-Deep data as determined from our simulations. At bright magnitudes ($H_{160,AB}<26$ mag) $>50$\% of all simulated $z=7.3-8.9$ galaxies are selected. At fainter magnitudes, this fraction is reduced due to photometric scatter in the color measurements and due to incompleteness.
  The selection function for CANDELS-Wide is essentially identical, however, shifted by 0.7 mag due to the difference in the WFC3/IR data.  
 \textit{ Bottom -- } The redshift distribution of the CANDELS $z\sim8$ galaxy candidates (blue thick line). The redshift distribution peaks at lower redshift than the selection function shown in the top panel due to the dimming of galaxies with redshift. The mean redshift of our sample is $\langle z \rangle = 7.9$, with 80\% of galaxies expected to lie at $z=7.2-8.7$. The best-fit LF determined in Section \ref{sec:z8LF} was used for determining this redshift distribution. Also shown are the redshift distribution functions of the HUDF09 (gray solid) and ERS (gray dashed) $z\sim8$ candidates from \citet{Bouwens11c}, which are very similar to the ones derived here for the CANDELS data.
  }
	\label{fig:selfun}
\end{figure}

\section{The $z\sim8$ LF}
\label{sec:LFconstraints}

We will now use the $z\sim8$ galaxy candidates identified in the previous section to derive constraints on the $z\sim8$ UV LF, and combine these
new results with previous estimates to give the best available LF at
$z\sim8$.

\subsection{Selection Functions and Redshift Distributions}

To compute the LF, we first have to estimate the completeness, $C(m)$, and redshift selection functions, $S(z,m)$. Following \citet{Oesch07,Oesch09}, this is done by inserting artificial galaxies with varying magnitudes, profiles and sizes in the observational data and rerunning the source detection with the exact same setup as for the original catalogs. This is done for each of the fields individually. 

Our simulations are based on using real galaxies at lower redshift and
scaling them to higher redshifts using well-established evolutionary and
cosmological relationships, i.e., using the 
`cloning' methodology of \citet{Bouwens03}. In particular, we `clone' $z\sim4$ LBGs from the GOODS and HUDF fields to higher redshifts.
 The images of these $z\sim4$ sources are scaled to the desired input magnitude, and are stretched to account for the difference in angular diameter distance, as well as a size scaling of $(1+z)^{-1}$ as observed for the Lyman Break galaxy population across $z\sim3-7$ \citep[see e.g.][]{Ferguson04,Bouwens04a,Oesch10b}. 
This procedure ensures that the distribution of morphologies and profiles of the
simulated galaxies is as close to reality as possible and thus specifically
accounts for SB-dimming effects, which result in loosing larger, resolved galaxies
from the samples.  
 The cloned galaxies are then inserted in the observed images with galaxy colors as expected for star-forming galaxies between $z=6$ and $z=9.5$ (see also Figure \ref{fig:colcol}). The adopted colors are based on a UV continuum slope distribution of $\beta=-2.5\pm0.4$   motivated by recent determinations of the UV continuum slopes as a function of UV luminosity at $z>6$ \citep[see e.g.][]{Bouwens09b,Bouwens10b,Stanway05,Finkelstein10,Finkelstein11,Wilkins11b,Dunlop11}. 
 As a cross-check we have also tested that our simulation pipeline returns essentially equivalent results when adopting theoretical Sersic galaxy profiles \citep[see also][]{Bouwens11c,Oesch11}.

From the simulation output, we compute the completeness as a function of observed $H_{160,AB}$ magnitude for each field, taking into account the scatter and bias between input and output magnitudes. Additionally, we compute the selection probabilities as a function of redshift and magnitude by measuring the fraction of sources that meet our selection criteria.  The selection function and the corresponding redshift distribution function is shown in Figure \ref{fig:selfun}. As can be seen, galaxies in our sample are selected from $z\sim7.2-8.7$, with a mean redshift of $\langle z \rangle = 7.9$.

We note that the expected redshift distribution has wings which extend to lower and higher
redshifts, and we expect $\sim4$ objects to lie at $z<7.5$. The shape of the selection
function has to be taken into account when modeling the observed number
densities, particularly if there is strong evolution over the redshift range
$z=7-8$. The redshift selection function is available from the authors upon
request.

\begin{figure}[tbp]
	\centering
	\includegraphics[scale=0.5]{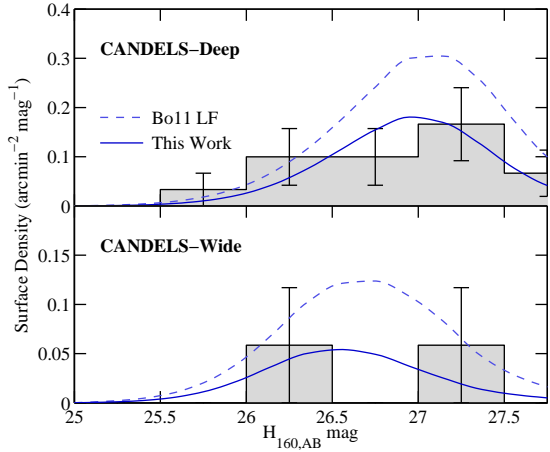} 
  \caption{The observed and the expected surface density of $z\sim8$ galaxies in GOODS-South. The two panels show the results for the CANDELS-Deep (upper) and CANDELS-Wide (lower) fields. The gray histograms show the observed surface density of candidates, while the two dark blue lines represent the expectations derived from our simulations. The dashed line is derived from the best-fit $z\sim8$ LF of \citet{Bouwens11c}. This somewhat overpredicts the observed number of sources at bright magnitudes found in our larger area survey. The solid line shows the expected surface density using the new best-fit LF as derived in Section \ref{sec:z8LF}. The surface densities peak before the 5 sigma detection limits because of the
reduction in the selection volume due to photometric scatter.}
	\label{fig:Nexp}
\end{figure}

\subsection{Expected Surface Density of $z\sim8$ Galaxies}
\label{sec:Nexp}

It is very instructive to compare the expected surface density of $z\sim8$ galaxy candidates using the previous best-fit LF of \citet{Bouwens11c} with the observed number of candidates in the CANDELS data. The expected number of sources in a given magnitude bin $m_i$ can be estimated for any given LF, $\phi(M)$, through:
\[
N^\mathrm{exp}_i  = \int_{\Delta m} dm \int dz \frac{dV}{dz} S(m,z)C(m)\phi(M[m,z])
\]

In Figure \ref{fig:Nexp} we show the histograms of the observed surface density of sources in the different fields and compare them to the expectation from different LFs. As can be seen, the expected source density peaks around 0.2 arcmin$^{-2}$ mag$^{-1}$ in the CANDELS-Deep field. The peak occurs   at $H_{160,AB}=27$ mag, which is $\sim$0.5 mag brighter than the formal 5$\sigma$ limiting magnitude. This is mainly due to the reduction in the selection volume due to photometric scatter, which causes us to lose sources from the color-color selection window near the limit.
Note that we do also expect a small fraction of sources even below the formal magnitude limit due to scatter in the photometric offsets between the aperture fluxes (from which the selection S/N is computed) and the total fluxes (used in Figure
\ref{fig:Nexp}), as well as from variations in the depth of the $H_{160}$ band data.

From the integration of the surface density expected from the best-fit
LF of \citet{Bouwens11c}, we find that we would expect to detect 22 $z\sim8$ galaxies in the CANDELS-Deep field, and five in CANDELS-Wide. This is a factor $1.7\times$ higher than the 16 candidates we find, indicating that the bright end of the previous UV LF might have been estimated somewhat high due to the presence of a few very bright sources ($H_{160,AB} \sim 26$ mag) in a possible overdensity in the HUDF09-2 field \citep[see also discussion in ][]{Bouwens11c}. The  difference
seems large but it is driven by just a few bright sources and so is
consistent with small number statistics.

\begin{figure}[tbp]
	\centering
	\includegraphics[width=\linewidth]{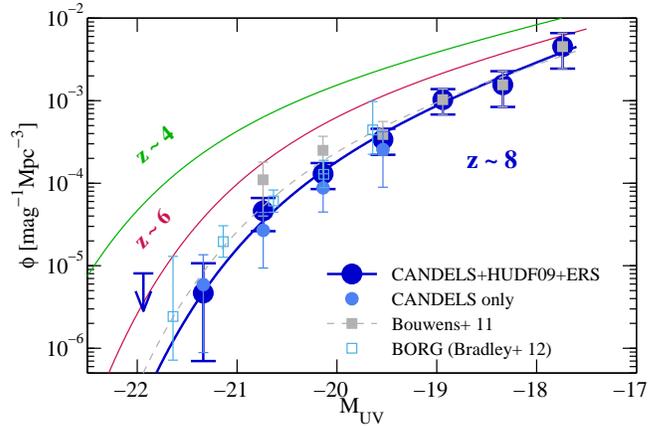} 
  \caption{The $z\sim8$ UV LF from the total WFC3/IR data in GOODS-South. The stepwise LF from the CANDELS data alone is plotted as small bright blue circles, while the previous determination from \citet{Bouwens11c} using the ultra-deep HUDF09 and the ERS data is shown as light gray squares. The new CANDELS LF is clearly lower than the previous determination from the HUDF09+ERS at all magnitudes $M_{UV}<-20$ by a factor $\sim3-4$.
  The best-fit LF from the HUDF09 and ERS data alone is shown as dashed gray line. The combined stepwise LF is shown as large blue circles, along with the best-fit as the solid blue line. The upper limit at the bright end corresponds to a $1\sigma$ limit for a non-detection.
  This new LF was obtained by combining all the 75 $z\sim8$ candidates from the current CANDELS, HUDF09 and ERS data (see section \ref{sec:z8LF}). It is also listed in Table \ref{tab:z8lf}. For comparison, we also show the LF determination from the BORG survey as open, light blue squares \citet{Bradley12}, which is in good agreement with our total LF determination.  }
	\label{fig:z8LF}
\end{figure}

\begin{figure*}[tbp]
	\centering
	\includegraphics[scale=0.7]{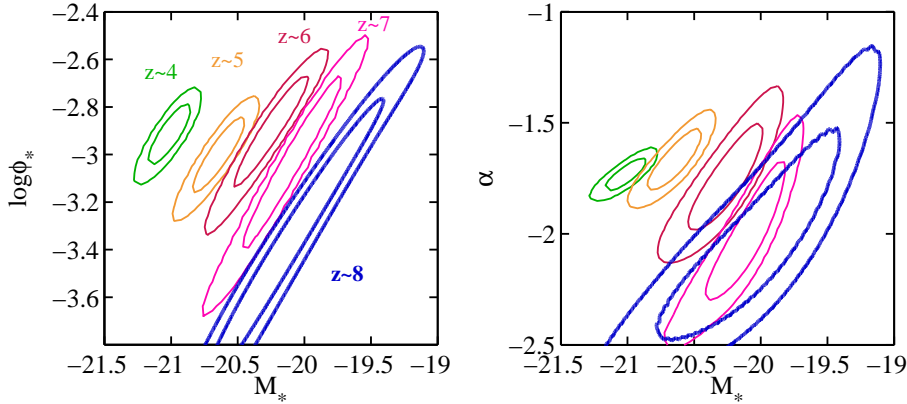} 
  \caption{The evolution of the Schechter function parameters and their uncertainties as a function of redshift. The contours show 68\% and 95\% of the likelihood at each redshift bin. The $z<8$ contours are from \citet{Bouwens11c}. Despite the large uncertainties, the $z\sim8$ LF parameters are significantly different from $z\sim7$. The difference is significant at $>99\%$. The characteristic magnitude is fainter by about 0.1 mag at $z\sim8$, consistent with pure luminosity evolution, i.e. a dimming of $M_*$ to higher redshifts. The faint-end slope $\alpha$ is very
steep, but it is still very uncertain at $z\sim8$. Future deeper data will be required to improve on this (for reionization) crucial measurement of
the slope $\alpha$.  }
	\label{fig:contours}
\end{figure*}

\subsection{New Constraints on the $z\sim8$ LF}

We now use the new CANDELS $z\sim8$ candidates to derive direct constraints on the UV LF, by computing the step-wise LF in bins of absolute magnitudes.
This is done using an approximation of the effective selection volume as a function of observed magnitude $V_{\rm eff}(m) = \int_0^\infty dz \frac{dV}{dz} S(z,m)C(m)$. The LF is then given by $\phi(M_i)dM = N^{\rm obs}_i/V_{\rm eff}(m_i)$. This is shown in Figure \ref{fig:z8LF}, where we evaluated the LF in bins of 0.6 mag, as a compromise between luminosity resolution and S/N. The error bars include a $30\%$ ($40\%$) contribution from cosmic variance for CANDELS-Deep (CANDELS-Wide), which we estimate using the cosmic variance calculator of \citet{Trenti08}; see also \citet{Robertson10a}.

As expected from the low observed surface density of $z\sim8$ candidates in the CANDELS data, the step-wise LF is significantly below the previous best-fit Schechter function from \citet{Bouwens11c}. At $M_{UV} = -21$ to $-19$ mag, we find values $\sim2-4\times$ lower than from the HUDF09 and ERS data. As noted above, these are still consistent given the current and previous uncertainties. Nevertheless, the CANDELS data indicates that the characteristic magnitude might be fainter than previously determined. We quantify this more precisely in the next section.

Finally, we also compare our LF determination with the bright end constraints from the BORG survey \citep{Bradley12}, which are based on 33 $z\sim8$ galaxy candidates with $J_{125}<27.4$ mag identified over an effective search area of  274 arcmin$^2$ (i.e. about $1.7\times$ the area of the HUDF09+ERS+CANDELS). As Figure \ref{fig:z8LF} shows, also the BORG $z\sim8$ LF supports the indication that the HUDF09 field is somewhat overdense in bright $z\sim8$ galaxies. The LF from BORG is a factor $\sim2-3\times$ lower than the step-wise LF determination from the HUDF09+ERS alone, and it is in excellent agreement with our new, total step-wise LF.

\begin{deluxetable}{cc}
\tablecaption{Stepwise Determination of the $z\sim8$ UV LF Based on CANDELS, HUDF09, and ERS Candidates \label{tab:z8lf}}
\tablewidth{210 pt}
\tablecolumns{2}
\tablehead{$M_{UV}$ [mag] & $\phi_*$  [10$^{-3}$Mpc$^{-3}$mag$^{-1}$]   }

\startdata

$  -21.94$ &   $<0.008$\tablenotemark{*} \\
$  -21.34$ &    0.005 $\pm$    0.006 \\ 
$  -20.74$ &    0.046 $\pm$    0.020 \\ 
$  -20.14$ &    0.130 $\pm$    0.045 \\ 
$  -19.54$ &    0.339 $\pm$    0.118 \\ 
$  -18.94$ &    1.03 $\pm$    0.35 \\ 
$  -18.34$ &    1.56 $\pm$    0.72 \\ 
$  -17.74$ &    4.52 $\pm$    2.07 

\enddata

\tablenotetext{*}{1$\sigma$ upper limit for a non-detection.}

\end{deluxetable}

\begin{deluxetable*}{lcccccccc}
\tablecaption{Comparison of $z\sim8$ LF Determinations in the Literature \label{tab:lfcomparison}}
\tablewidth{0 pt}
\tablecolumns{4}
\tabletypesize{\scriptsize}
\tablehead{Reference & $\log\phi_*$  [Mpc$^{-3}$mag$^{-1}$]  &  $M_{UV}^*$ [mag]  &  $\alpha$  }

\startdata
This Work  				&  $-3.30^{+0.38}_{-0.46}$  & $-20.04 ^{+0.44}_{-0.48}$  & $-2.06^{+0.35}_{-0.28} $  \\  
\citet{Bradley12}		&  $-3.37^{+0.35}_{-0.21}$   & $-20.26^{+0.29}_{-0.34}$  &  $-1.98^{+0.23}_{-0.22}$ \\
\citet{Bouwens11c}  	&  $-3.23^{+0.74}_{-0.27}$ & $-20.10\pm0.52$  &  $-1.91\pm0.32$   \\
\citet{Lorenzoni11}	&  $-3.0$  & $-19.5$  &  $-1.7$ (fixed) \\
\citet{Trenti11a} 	    &  $-3.4$ (fixed) & $ -20.2\pm0.3$  & $-2.0$ (fixed)\\
\citet{McLure10}		&  $-3.46$ & $-20.04$ (fixed)  &  $-1.71$ (fixed) \\
\citet{Bouwens10a}	&  $-2.96$ (fixed)  & $-19.5\pm0.3$  & $-1.74$ (fixed)

\enddata

\end{deluxetable*}

\subsection{Combination with Deeper Data: The $z\sim8$ Schechter Function}
\label{sec:z8LF}

Due to the small dynamic range in luminosities, the bright candidates identified in the CANDELS data are not sufficient to provide a good estimate of the overall shape of the UV LF alone. We therefore combine our new sources with all the $z\sim8$ candidates from ultra-deep field measurements from \citet{Bouwens11c} to update the Schechter function parameters of the $z\sim8$ UV LF. By doing so, we can now generate a LF using a total of 75 $z\sim8$ candidates identified over 148 arcmin$^2$, spanning $H_{160,AB} \sim 25.7-29.5$ mag. The stepwise determination of this combined sample is listed in Table \ref{tab:z8lf} and shown in Figure \ref{fig:z8LF}.

The Schechter function parameters are derived by maximizing the Poissonian likelihood for observing $N^{\rm obs}$ sources in a given magnitude bin when $N^{\rm exp}$ are expected to be seen based on a given UV LF. We thus maximize  $\cal{L} $ $= \prod_{j} \prod_i P(N^{\rm obs}_{j,i},N^{\rm exp}_{j,i})$, where $j$ runs over all fields, and $i$ runs over the different magnitude bins, and $P$ is the Poissonian probability.
The expected number of sources are computed according to the equation from section \ref{sec:Nexp}. For the HUDF09 and ERS fields we adopt the selection volumes estimated in \citet{Bouwens11c}. The combination of these older
results with our new sample is appropriate since both are based on essentially identical simulations and inputs, i.e., our estimates of the CANDELS selection functions match those of \citet{Bouwens11c}.

The best-fit parameters are determined by a grid search over Schechter function parameters, maximizing the combined likelihood $\cal{L}$. The best-fit solutions are: $\log(\phi_*$ [Mpc$^{-3}$mag$^{-1}])=-3.30^{+0.38}_{-0.46}$, $M_* = -20.04 ^{+0.44}_{-0.48}$ mag, and  $\alpha = -2.06^{+0.35}_{-0.28}$. 
These parameters are consistent with the previous determination from the HUDF09 and ERS data alone \citep{Bouwens11c}, and, despite a somewhat larger sample size, the uncertainties are not markedly reduced. However, the best-fit characteristic magnitude is fainter by $\sim0.1$ mag, which is due to  the lower number of detected sources in the CANDELS data than the number that was expected from the previous LF determination.

Note also that the faint-end slope is steeper than $\alpha=-2$, as discussed
by \citet{Bouwens11b}. This leads to a formally divergent luminosity density. However, galaxies are not expected to be formed below a given luminosity due to inefficient cooling in low mass halos and feedback effects (at masses that correspond to $M_{UV}$ about $-$10 to $-$11), and so
the luminosity density converges. Nevertheless, such
steep slopes have important consequences for reionization by galaxies as pointed out previously \citep[e.g.][]{Bouwens11b}.

At the bright end ($M_{UV}<-20$), our best-fit LF is a factor $\sim1.7\times$ lower than the previous determination of \citet{Bouwens11c}, who used only the HUDF09 and ERS data.   
However, \citet{Bouwens11c} only find such a higher surface density for $z\sim8$
sources in the HUDF09, while the number counts over the ERS are actually in excellent agreement with what we find over the CANDELS field.
Therefore, all the wide-area data over the CDF-South GOODS yield
approximately the same surface density of bright $z\sim8$ galaxies. 
Nevertheless, we stress again that both the step-wise and the best-fit LFs are consistent with the previous determination of \citet{Bouwens11c}, given the still small number of sources at the bright end.


In Table \ref{tab:lfcomparison}, we compare our new best-fit LF parameters with other, previous determinations from the literature \citep[][]{Bouwens10a,Bouwens11c,McLure10,Lorenzoni11,Trenti11a,Bradley12}. Within the current measurement uncertainties these are all consistent with each other. 

Overall, it is very reassuring that different groups arrive at similar results given the variety of approaches.  Nonetheless, it is worthwhile remarking that with the exception of \citet{Bouwens11c} and \citet{Bradley12} the previous determinations were not based on large enough data sets such that all three Schechter function parameters could reliably be fit simultaneously as done here.

\subsection{Evolution of the LBG Population at $4<z<8$}

A key diagnostic in studying the build-up of galaxies is how the UV LF evolves with redshift, as this is directly related to the distribution of SFRs in galaxies. In Figure \ref{fig:contours}, we show the error
contours of the Schechter function parameters from current \textit{HST} data of LBGs at different redshifts. The error contours for the $4\leq z \leq 7$ samples are taken from \citet{Bouwens07,Bouwens11c}, while the $z\sim8$ contours correspond to the likelihood contours of our Schechter function fit at $z\sim8$ from all available data, i.e. including the CANDELS GOODS-South candidates (Section \ref{sec:z8LF}). 

The Schechter function parameters show significant evolution from $z\sim7$ to $z\sim8$. In particular, the combination of the characteristic magnitude and number density is evolving at $>99\%$ significance. 
On the other hand, the combined constraint on the faint-end slope and the characteristic luminosity is still quite weak (right panel of Figure \ref{fig:contours}). Unfortunately, the faint-end slope is largely unconstrained. This is very unfortunate since it is one of the most important parameters for assessing the contribution of galaxies to reionization. The flux density of ionizing photons from galaxies is extremely sensitive to the faint-end slope $\alpha$, especially when the slope is as
steep as $\alpha \sim -2$ \citep[e.g.][]{Bouwens11b}. A continued effort in the future will thus be to constrain this parameter better with deeper WFC3/IR imaging.

\begin{figure}[tbp]
	\centering
	\includegraphics[width=\linewidth]{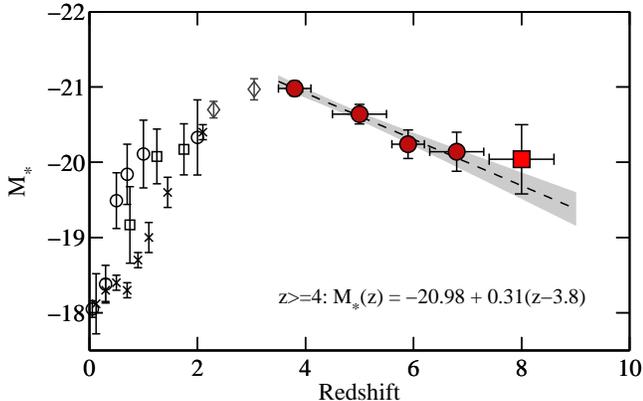}  
  \caption{Evolution of the characteristic luminosity of the UV LF across redshift. The measurements at $z>3.5$ shown as dark red circles are based on LBG selections in the deepest HST images, including the HUDF and HUDF09 fields. These are taken from \citet{Bouwens07,Bouwens11c}. In general, determinations by other authors are in good agreement with these measurements.  
  The red square is the best-fit value for M$_*$ at $z\sim8$ when combining the HUDF09+ERS data with our new measurements from the CANDELS data (see section \ref{sec:z8LF}). The lower redshift measurements are a selection of UV LF parameters determined by \citet[][gray diamonds]{Reddy09} at $z\sim2$ - 3, and \citet[][black open squares]{Oesch10d}, \citet[][black open circles]{Arnouts05}, and \citet[][black crosses]{Cucciati11} at $z<2$. The dashed black line and gray shaded area correspond to the best-fit evolution of $M_*$ as a function of redshift at $z>4$.  }
	\label{fig:MstarEvol}
\end{figure}

Overall, the UV LF constraints of LBGs are consistent with pure luminosity evolution from $z\sim8$ to $z\sim4$. This can easily be achieved by growing SFRs of individual galaxies \citep[e.g.][]{Stark09,Finlator11a,Papovich11,Smit12,Jaacks12}. The best-fit evolution of $M_*$ of LBGs at $z>3$ follows: 
\[
M_*(z) = -20.98(\pm0.04) + 0.31(\pm0.03)\times(z-3.8).
\]
 This is shown in Figure \ref{fig:MstarEvol}, where we plot the evolution of the characteristic cut-off luminosity of the UV LF with redshift. After correcting for dust extinction, the SFR of an $L_*$ galaxy thus grows by almost an order of magnitude from $z\sim8$ to $z\sim4$.

The value of $M_*$ as measured at $z\sim8$ is essentially equal to the cut-off UV luminosity at $z\sim1.5$-$2$, just shortly after the peak of the cosmic SFR density. However, the SFR of an $L_*$ galaxy is nevertheless larger at $z\sim1.5$-$2$ than at $z\sim8$ due to the larger dust obscuration \citep[e.g.][]{Reddy10,Bouwens11d,Smit12}.

\begin{figure}[tbp]
	\centering
	\includegraphics[width=\linewidth]{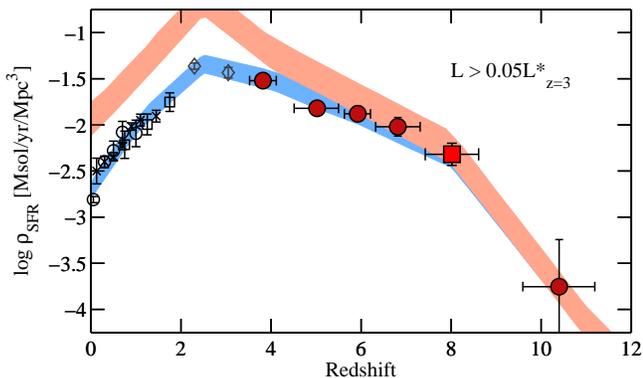}  
  \caption{The evolution of the star-formation rate density with redshift. The measurements are based on the integration of the UV LFs to $M_{UV}=-17.7$  and converting to SFRs using the relation of \citet{Madau98}. The upper, red shaded area is obtained after correcting the observed UV luminosity densities for dust extinction based on the UV continuum slope distributions of \citet{Bouwens11d}.
  The plot symbols are the same as in  Figure \ref{fig:MstarEvol}. The SFR density estimate at $z\sim10$ is based on the detection of one galaxy candidate in the HUDF \citep{Oesch11}. The rapid (but still uncertain) increase in the cosmic SFR density from $z\sim10$ to $z\sim8$ will be refined with upcoming WFC3/IR imaging over the HUDF. }
	\label{fig:sfrd}
\end{figure}

In Figure \ref{fig:sfrd} we show how the (dust-corrected) SFR density evolves across redshift when including our new UV LF parameters and the updated dust corrections for LBGs at $z>4$ from \citet{Bouwens11d}. The SFR density includes all galaxies down to a fixed flux limit of $M_{UV}=-17.7$, which is the current detection limit at $z\sim8$. Our updated SFR density at $z\sim8$ is only marginally lower than the previous measurement of \citet{Bouwens11c} by 0.07 dex, i.e. we derive $\log\rho_{SFR} = -2.32\pm0.12$ $M_\odot $yr$^{-1}$Mpc$^{-3}$. This is based on the luminosity density of $\log\rho_{L} = 25.58\pm0.12$ erg~s$^{-1}$Hz$^{-1}$Mpc$^{-3}$ and the conversion of the UV luminosity to SFR by \citet{Madau98}. 

As can be appreciated from Figure \ref{fig:sfrd}, the SFR density grows rather dramatically by more than an order of magnitude from $z\sim10$ \citep{Oesch11,Bouwens11a} to our new determination at $z\sim8$. After that it evolves very steadily, growing by another $\sim1.5$ dex from $z\sim8$ to its peak at $z\sim2.5$. The rapid growth at $z>8$ is very intriguing but still very uncertain. The change at
$z>8$ may well be refined with planned deep F140W data over the HUDF.

\section{Summary}
\label{sec:summary}

In this paper, we analyzed newly acquired WFC3/IR data over the GOODS-South field as part of the CANDELS MCT program to derive new constraints on the UV LF at $z\sim8$, about 600-750 Myr after the Big Bang. Unlike at $z\sim7$, where the brightest galaxies can be detected from the ground, $z\sim8$ galaxies are too faint to be detected with any
reliability with current ground-based instrumentation. Therefore, the bright end of the $z\sim8$ UV LF has remained relatively uncertain since it has been based on WFC3/IR data that was only available over a relatively small area, the very deep HUDF09 fields.

Galaxies at $z\sim8$ are identified using the Lyman Break technique on optical and near-IR data, and particularly the $Y_{105}$ band,
leading to them being identified as `$Y_{105}$-dropouts'. The CANDELS fields over GOODS are extremely valuable data sets, as they combine imaging in three WFC3/IR filters, and in addition also have relatively deep ancillary optical data from \textit{HST} ACS, which allows for such $z\sim8$ galaxy selections. Our selection criteria are outlined in section \ref{sec:selection} and were chosen to match prior measurements in the
ultra-deep HUDF09 dataset. They select galaxies at $z\sim7.2-8.7$ with a mean redshift $\langle z \rangle = 7.9$.

In the full search area of $\sim95$ arcmin$^2$, we identified 16 new $z\sim8$ galaxy candidates with $H_{160,AB}$ magnitudes in the range $25.7-27.9$ mag. These are presented in Figure \ref{fig:stamps} and Table \ref{tab:phot}. The $z\sim8$ candidates were selected using very strict optical non-detection requirements, including a measurement of the optical $\chi^2_{opt}$ flux (see Section\ref{sec:chi2opt}). This allows us to reduce the expected contamination due to photometric scatter and due to the limited depth of the optical data relative to the WFC3/IR imaging by a factor $\sim3-4$. Our sample is thus expected to show relatively low levels of contamination, which we estimate to be $\sim10\%$.

Interestingly, the observed surface density of $z\sim8$ galaxies is $\sim1.7\times$ lower than expected from the previous best-fit $z\sim8$ LF of \citet{Bouwens11c}, though the difference is consistent with the statistics
from the small numbers of objects at bright magnitudes in the previous
search. This previous estimate was based on the much smaller area data from the HUDF09 fields and the ERS data. By using the current CANDELS imaging we triple the search volume for luminous galaxies at $M_{UV}<-19.5$ mag in the CDFS relative to our previous analysis, thus reducing the potential biases introduced by cosmic variance. 

We combine our new CANDELS $z\sim8$ candidates with our previous candidates from the much deeper HUDF09 and ERS fields to derive the best possible measurement of the UV LF.
The best-fit $z\sim8$ UV LF we derive in this way is consistent with the previous estimates from \citet{Bouwens11c}. However, the best-fit characteristic magnitude is fainter by $\sim0.1$ mag ($M_{*}^{z=8} = -20.04\pm0.46$ mag). Despite the larger area probed here, the uncertainties on the Schechter function parameters are still significant. Nonetheless,
these new results are contributing to our growing understanding of the
evolution of key parameters in the luminosity function from $z\sim8$ to $z\sim2$, and
of the star formation rate density from $z\sim10$ to $z\sim2$.

 A combination of future wide area data, as well as deeper WFC3/IR imaging will be necessary to further improve on these very important measurements that are crucial for refining our estimates of the role of
UV photons from galaxies in the reionization of the universe before the advent of JWST.


\acknowledgments{We are grateful to Massimo Stiavelli, Kristian Finlator, Steve Finkelstein and Naveen Reddy for helpful discussions related to this work.
We also thank the CANDELS collaboration for their big efforts in planning and obtaining such an exquisite data set available to the community.
Support for this work was provided by NASA through Hubble Fellowship grant HF-51278.01. This work has further been supported by NASA grant HST-GO-11563.01. This research has benefited from the SpeX Prism Spectral Libraries, maintained by Adam Burgasser at http://www.browndwarfs.org/spexprism. }

Facilities: \facility{HST(ACS/WFC3), Spitzer(IRAC)}.


\appendix

\section{Comparison to Yan et al. (2011b) Catalog}

At the same time our paper was submitted, the CANDELS team submitted a very similar $z\sim8$ galaxy search in \citet{Yan12a}. 
That paper is still under review and is likely to change further (Yan 2012, private communication). Below we explicitly compare only to version 2 of their manuscript, which is publicly available on arXiv at the time of publication of our paper.
The \citet{Yan12a} analysis is not based on the complete GOODS-South CANDELS data set (they include 83\% of the final data in $J_{125}$ and $H_{160}$ that was used in our analysis). Additionally, they only analyze the CANDELS-DEEP area (see Fig \ref{fig:fields}), and they only include 8-epoch mosaics of the $I_{814}$ image. 

For comparison, our analysis is based on \textbf{all} the $I_{814}$ data available over the GOODS-South, which was obtained from a large number of additional programs (ERS, HUDF09, CANDELS SNe follow-up, 3D-HST, and the UVUDF). As noted in the main text, these data reach almost 1 mag deeper than the GOODS $i_{775}$ data, and are thus an extremely valuable addition to check for low-redshift contaminants.
Additionally, we include a deeper reduction of the other ACS images including additional data taken after the original GOODS-South program, which reaches to $\sim0.1-0.2$ mag deeper compared to the GOODS v2.0 data.

The revised v2 manuscript of \citet{Yan12a} lists a total of 16 different $Y_{105}$-dropout candidates, which are split in two different samples, one using AUTO fluxes and one using ISO fluxes for computing colors. Out of these 16 sources, only 5 are in common with our sample. \citet{Yan12a} use somewhat more restrictive color criteria than what we adopt here ($Y_{105}-J_{125}>0.8$, instead of $>0.45$ as used in our analysis). Therefore, we would have expected to select all their sources in our catalog.

Upon inspection of their sources, we found the main reason that \citet{Yan12a} $z\sim8$ candidates did not appear in our catalog is due to a large fraction of their candidates showing non-negligible flux in the optical ACS data such that they do not satisfy our strict optical non-detection criterion $\chi^2_{opt}<3$.
 This applied to 7 out of their 16 candidates. 
\citet{Yan12a} state that they perform a two-stage source selection, which includes a visual inspection of their candidates using all the data. However, some sources show detections in our reduction of the optical ACS data,  disqualifying these sources being at $z>7$. Furthermore, two sources are detected in the HUDF ACS data. Since these data were used in their visual inspection, we do not have an explanation for this discrepancy.


In addition to these 7 sources with optical flux, their list also includes two sources which are not significantly detected ($<5\sigma$) in our $H_{160}$-band image, as well as two sources for which we measure $Y-J$ colors that are too blue, although they are consistent with being $Y$-dropouts at the $1\sigma$ level. 

\begin{figure}[tbp]
	\centering
	\includegraphics[scale=0.55]{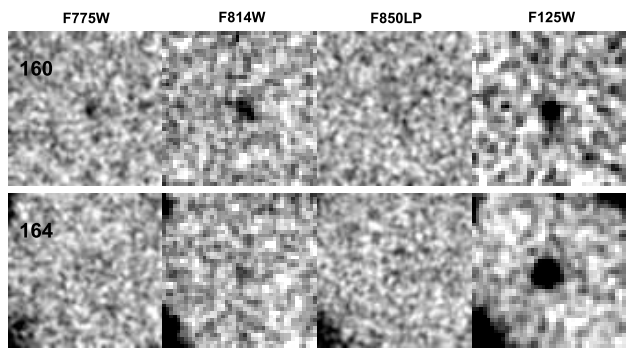}
  \caption{Two examples of clear optical detections in the \citet{Yan12a} sample (ISO\_160, and ISO\_164). Both sources show a detection in $I_{814}$, and show hints of flux in other bands as well. The increased depth of the $I_{814}$ data is clearly an extremely valuable addition for removing contaminants. Note that both these sources were flagged as contaminants in our catalog based on our $\chi^2_\mathrm{opt}<3$ criterion, showing the power of using this measurement.   }
	\label{fig:yanYdrops}
\end{figure}

A detailed summary of all our measurements and analysis of the \citet{Yan12a} sample is provided in Table \ref{tab:yancomparison}, and an example of optical stamps for two sources are shown in Figure \ref{fig:yanYdrops}.

Given this disagreement, it is encouraging, however, that \citet{Yan12a} do, in fact, include 5 out of the 6 sources with $S/N(H_{160}) > 7$ from our sample which satisfy their stricter color criterion of $Y-J>0.8$ (see Table \ref{tab:phot}). This indicates that our selection is quite robust, at least for higher-significance sources, if optically detected interlopers are excluded properly, as done for our catalog.

We stress again that making full use of all the information in the optical data is extremely important for a reliable LBG selection. We did our best possible effort to do this by including a limit in the optical $\chi^2_\mathrm{opt}$ measurement and by analyzing all the available optical ACS data that were taken over this field. Furthermore, we note that as long as the contamination fractions and detection efficiencies are modeled self-consistently, using shallower data should not result in a different estimate of the final LF (within the errors).



\begin{deluxetable}{lccccccl}
\tabletypesize{\scriptsize}
\tablecaption{Comparison to Source List of Yan et al. (ApJ submitted, arXiv:1112.6406v2) \label{tab:yancomparison} }
\tablehead{\colhead{Yan ID} & $\alpha$ & $\delta$ &\colhead{$J_{125}$} &\colhead{S/N$_H$} &\colhead{$J_{125}-H_{160}$} & \colhead{$Y_{105}-J_{125}$} & \colhead{Note}  }

\startdata
ISO\_085  &  03:32:49.94  &  -27:48:18.1  &  25.8 $\pm$ 0.1 (25.6)  & 22 & 0.02 $\pm$ 0.08 ($-$0.1) & 1.00$\pm$0.11 (1.20) & OK, in our catalog  \\ 
ISO\_164  &  03:32:41.42  &  -27:44:37.8  &  26.1 $\pm$ 0.1 (26.0)  & 17 & $-$0.00 $\pm$ 0.10 (0.0) & 0.82$\pm$0.14 (0.90) & $\chi^2_{opt} = 3.3 > 3$  \\ 
ISO\_157  &  03:32:42.88  &  -27:45:04.3  &  26.8 $\pm$ 0.1 (26.5)  & 10 & 0.07 $\pm$ 0.18 (0.1) & 0.76$\pm$0.27 (0.90) & $\chi^2_{opt} =8.7 > 3$  \\ 
ISO\_071  &  03:32:20.98  &  -27:48:53.5  &  27.1 $\pm$ 0.2 (26.9)  & 8.2 & 0.07 $\pm$ 0.23 (0.1) & 1.32$\pm$0.48 ($>$1.70) & OK, in our catalog  \\ 
ISO\_082  &  03:32:14.13  &  -27:48:28.9  &  26.9 $\pm$ 0.1 (27.2)  & 7.1 & $-$0.47 $\pm$ 0.24  (0.0) & $>$2.1 ($>$1.10) & S/N(I$_{814})=2.1$  \\ 
ISO\_078  &  03:32:41.65  &  -27:48:34.5  &  27.4 $\pm$ 0.2 (27.2)  & 5.0 & $-$0.01 $\pm$ 0.26  (0.2) & $>$2.3 ($>$1.70) & $\chi^2_{opt} = 4.6 > 3$\tablenotemark{a}  \\ 
ISO\_011  &  03:32:14.47  &  -27:51:48.5  &  27.4 $\pm$ 0.3 (27.2)  & 3.6 & $-$0.35 $\pm$ 0.39  (0.0) & $>$1.7 (1.20) & too low S/N in $H_{160}$ \\ 
ISO\_158  &  03:32:47.95  &  -27:44:50.4  &  27.5 $\pm$ 0.2 (27.3)  & 7.6 & 0.34 $\pm$ 0.24 (0.2) & 0.39$\pm$0.30 (0.80) & too blue $Y-J$  \\ 
ISO\_063  &  03:32:35.00  &  -27:49:21.6  &  27.3 $\pm$ 0.2 (27.6)  & 7.1 & 0.10 $\pm$ 0.22  (0.1) & $>$2.3 (1.50) & OK, in our catalog  \\ 
ISO\_017  &  03:32:18.09  &  -27:51:18.5  &  27.9 $\pm$ 0.2 (27.6)  & 4.0 & $-$0.04 $\pm$ 0.31  ($-$0.4) & $>$1.5 (1.30) & $\chi^2_{opt} =3.01 > 3$  \\ 
ISO\_160  &  03:32:46.11  &  -27:44:48.0  &  27.8 $\pm$ 0.4 (27.9)  & 3.8 & $-$0.07 $\pm$ 0.33 (0.1) & 0.20$\pm$0.30 (0.90) &  S/N$_{z}=2.8$, S/N$_{i}=3.1$\tablenotemark{b} \\ 
ISO\_008  &  03:32:16.91  &  -27:52:01.9  &  27.7 $\pm$ 0.3 (28.0)  & 2.7 & $-$0.75 $\pm$ 0.57  ($-$0.3) & $>$1.5 (0.90) & too low S/N in $H_{160}$  \\ 
AUTO\_212  &  03:32:20.96  &  -27:51:37.1  &  26.4 $\pm$ 0.1 (26.4)  & 7.8 & $-$0.19 $\pm$ 0.16 (0.1) & 1.15$\pm$0.26 (1.00) & OK, in our catalog  \\ 
AUTO\_368  &  03:32:34.50  &  -27:46:03.5  &  27.1 $\pm$ 0.2 (27.2)  & 7.5 & 0.30 $\pm$ 0.19 (0.2) & 0.61$\pm$0.33 (0.80) & S/N$_z=2.6$, S/N$_i=2.1$\tablenotemark{c}  \\ 
AUTO\_204  &  03:32:18.18  &  -27:52:45.6  &  27.3 $\pm$ 0.2 (27.4)  & 7.0 & $-$0.16 $\pm$ 0.23 (0.2) & 1.06$\pm$0.44 (0.90) & OK, in our catalog  \\ 
AUTO\_094  &  03:32:40.67  &  -27:45:11.6  &  27.5 $\pm$ 0.2 (27.4)  & 4.1 & $-$0.24 $\pm$ 0.34 (0.0) & 0.28$\pm$0.30 (0.90) & too blue $Y-J$

\enddata
\tablenotetext{*}{The values in this table are our own measurements. For comparison, we show the measurements from Yan et al. in parentheses. Limits are $1\sigma$.}
\tablenotetext{a}{Source ISO\_078 is detected in several optical filters of the HUDF ACS data.}
\tablenotetext{b}{Source ISO\_160 is detected in all optical bands of our reduction of the GOODS data at more than 1.5$\sigma$.}
\tablenotetext{c}{Source AUTO\_368 is additionally detected in the $V_{606}$ data of the HUDF.}
\end{deluxetable}

\begin{figure*}[tbp]
	\centering
	\includegraphics[scale=0.7]{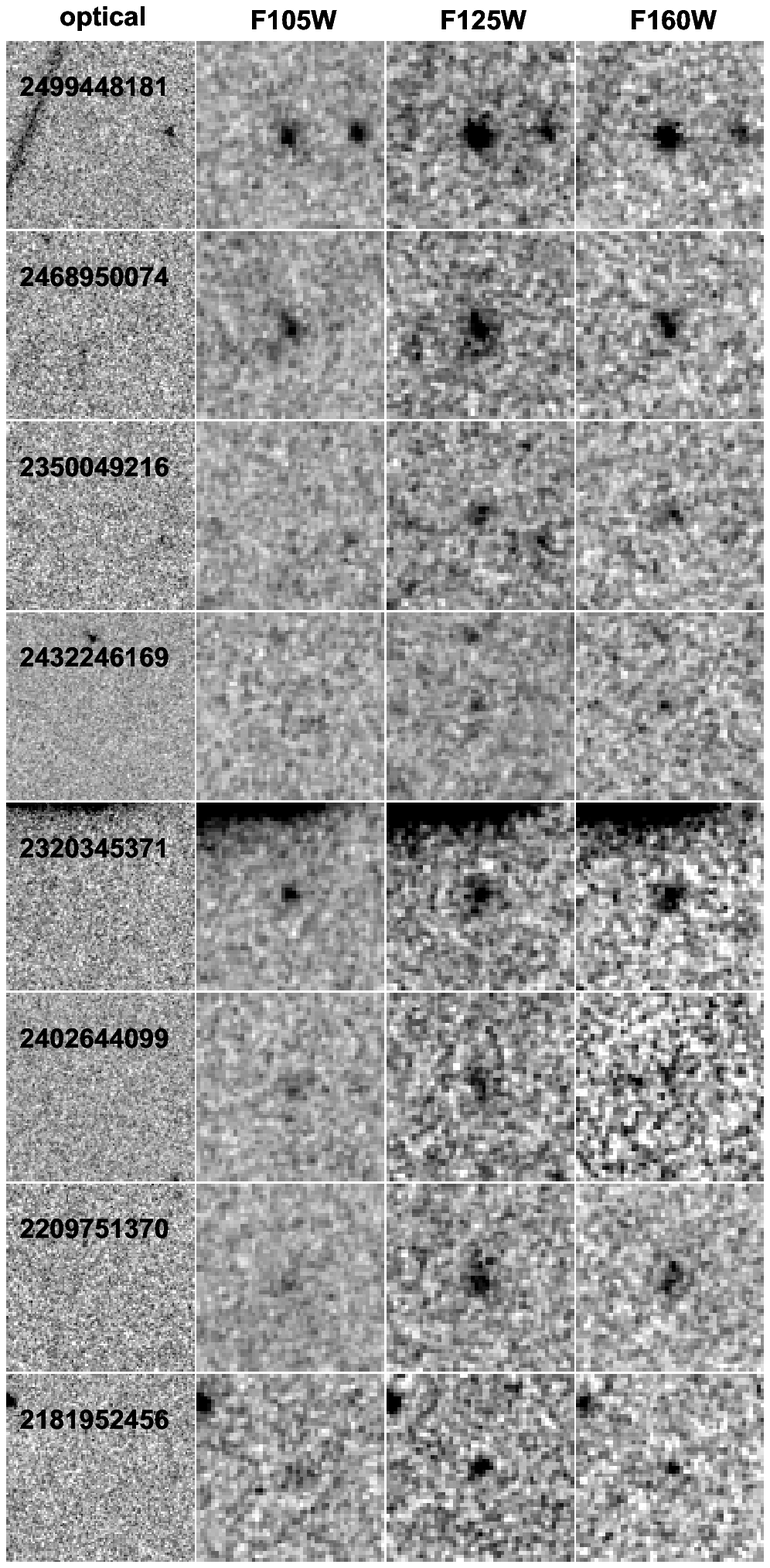}  
    \includegraphics[scale=0.7]{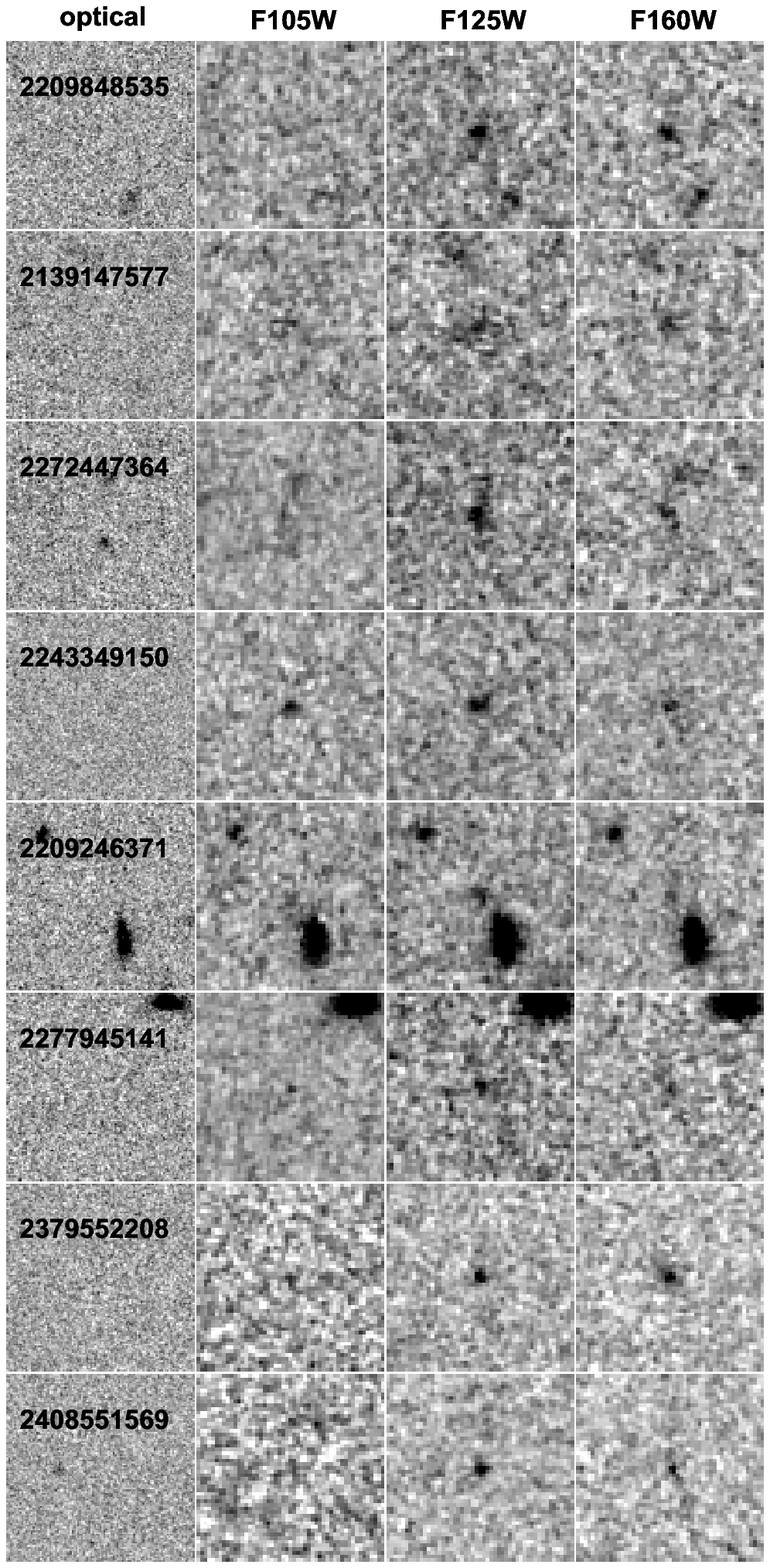} 
  \caption{ Same as Figure \ref{fig:stamps}, but without the slight smoothing of the stamps. Smoothing more accurately displays a larger dynamic range in the images and thus shows  the reliability of the sources more fairly. }
	\label{fig:stampssm}
\end{figure*}

\end{document}